\documentclass[12pt]{iopart}
\usepackage{cite}
\usepackage{geometry}
\usepackage[colorlinks=true,urlcolor=black,linkcolor=black,hyperfootnotes=false]{hyperref}
\usepackage{footnote}
\usepackage{graphicx,multirow}%
\usepackage{xcolor}
\usepackage{amsmath}
\usepackage{subcaption}	
\usepackage{cleveref}
\usepackage{setspace}
\usepackage{upgreek} 
\usepackage{array,longtable}
\newcounter{magicrownumbers}
\usepackage{multicol}
\usepackage{stackengine}
\usepackage{gensymb}

\usepackage{bm}
\usepackage[symbol]{footmisc}

\begin{document}

\title[Formation of O and H in an atmospheric pressure plasma containing humidity]{The formation of atomic oxygen and hydrogen in atmospheric pressure plasmas containing humidity: picosecond two-photon absorption laser induced fluorescence and numerical simulations}

\author{Sandra~Schr\"oter$^1$, J\'er\^ome~Bredin$^{1}$\footnote{Current affiliation: Space Science and Technology Department (RAL Space), STFC Rutherford Appleton Laboratory, Harwell Campus, Didcot, OX11 0QX, UK},
Andrew~R.~Gibson{$^{1,2,3}$}, Andrew~West{$^1$}\footnote{Current affiliation: Department of Electrical and Electronic Engineering, The University of Manchester, Manchester M13 9PL, UK}, James P.~Dedrick{$^1$}, Erik~Wagenaars{$^1$}, Kari~Niemi{$^1$}, Timo~Gans{$^1$}, Deborah~O'Connell{$^1$}}

\address{$^1$York Plasma Institute, Department of Physics, University of York, Heslington, York YO10 5DD, United Kingdom\\}
\address{$^2$Research Group for Biomedical Plasma Technology, Ruhr-Universität Bochum, Universitätsstraße 150, 44801 Bochum, Germany\\}
\address{$^3$Institute of Electrical Engineering and Plasma Technology, Ruhr-Universität Bochum, Universitätsstraße 150, 44801 Bochum, Germany}
\ead{deborah.oconnell@york.ac.uk}

\begin{abstract}
Atmospheric pressure plasmas are effective sources for reactive species, making them applicable for industrial and biomedical applications. We quantify ground-state densities of key species, atomic oxygen (O) and hydrogen (H), produced from admixtures of water vapour (up to 0.5\%) to the helium feed gas in a radio-frequency-driven plasma at atmospheric pressure. Absolute density measurements, using two-photon absorption laser induced fluorescence, require accurate effective excited state lifetimes. For atmospheric pressure plasmas, picosecond resolution is needed due to the rapid collisional de-excitation of excited states. These absolute O and H density measurements, at the nozzle of the plasma jet, are used to benchmark a plug-flow, 0D chemical kinetics model, for varying humidity content, to further investigate the main formation pathways of O and H. It is found that impurities can play a crucial role for the production of O at small molecular admixtures. Hence, for controllable reactive species production, purposely admixed molecules to the feed gas is recommended, as opposed to relying on ambient molecules. The controlled humidity content was also identified as an effective tailoring mechanism for the O/H ratio. 

\end{abstract}
\submitto{\PSST}

\section{Introduction}

Non-thermal atmospheric pressure plasma jets (APPJs) driven with radio-frequency (rf) power are very efficient sources of reactive species ~\cite{Graves2012,Lu2016,Dedrick2017,Wijaikhum2017,Murakami2014,Murakami2013,Niemi2013,Sousa2011,Maletic2012,Turner2016}. Reactive species play a crucial role in applications such as surface treatment \cite{Lommatzsch2007, Fang2013, Shaw2016}, etching \cite{Jeong1998, Fricke2011, West2016a} and biomedicine \cite{Laroussi2009, Park2012, Privat-Maldonado2016, Kong2009, Woedtke2013, Bekeschus2016}.
APPJs enable the localised delivery of reactive species to temperature sensitive biological samples~\cite{Morfill2009,Woedtke2013}. They have therefore generated considerable interest with respect to medical applications including wound healing~\cite{Tipa2011,Nastuta2011,Haertel2014,Isbary2012,Lloyd2010} and cancer therapies~\cite{Hirst2014, Hirst2015, Hirst2016, Gibson2014, Ratovitski2014, Mizuno2017, Vermeylen2016, Vandamme2011}. A key feature of APPJs is their potential to enhance treatment through the synergistic delivery of multiple reactive species, and other plasma components. To achieve optimised reactive species delivery, and treatment effectiveness, for a given application, it is crucial to understand the mechanisms behind the formation of important reactants and the chemical kinetics that occur both in the plasma itself and the plasma effluent, which is in direct contact with the treated sample.

Atomic species, such as atomic oxygen, hydrogen, and nitrogen (O, H, and N), are very reactive and are important precursors for longer lived species, such as nitrogen oxides N$_x$O$_y$, or ozone, which can play an important role in, for example, biomedical applications \cite{Graves2012}. Therefore, their precise quantification in the plasma effluent region is crucial in understanding underlying fundamental mechanisms, which can then help to optimise parameters such gas composition and treatment time. The quantification of reactive species in APPJs meets many challenges naturally arising from the geometry and characteristics of these sources when compared to low-pressure systems. Dimensions of APPJs are typically small, in the order of $\upmu$m to mm, requiring a high spatial resolution of the diagnostics that are used for the quantification of reactive species. On the other hand, the strong collisionality in APPJs can significantly reduce lifetimes of excited states in a radiation-less manner (quenching), posing additional challenges. 

An established diagnostic technique for quantifying atomic species such as O, H, and N is Two-photon Absorption Laser Induced Fluorescence (TALIF) \cite{Dobele2005, Zhang2014a, Gessel2013, Wagenaars2012, Schroder2012, Haensch1975, Bischel1981, Bokor1981,Kulatilaka2012}. TALIF measurements can provide high spatial resolution, in contrast to other techniques, such as for example absorption spectroscopy, which is typically used to measure line-of-sight averaged densities. TALIF is based on the measurement of fluorescence emission from a laser-excited state, which depends on all de-population mechanisms of that particular state, such as radiative de-excitation and radiation-less collisional quenching. Particularly the latter process can lead to a strong reduction of both the absolute fluorescence signal and the lifetime of the laser excited state at atmospheric pressure. Most conventional TALIF systems used for the investigations of APPJs comprise lasers and detection systems that operate with timescales in the region of nanoseconds, and are therefore not able to temporally resolve the excited state lifetime at elevated pressures. Although this can be calculated using quenching coefficients from the literature, uncertainties can be introduced due to the uncertainties associated with the rate coefficients for these processes, or because the gas mixture is complex and the distribution of quenching partners unknown. The latter is particularly the case for plasma effluents of APPJs, where a gradual mixing of the feed gas with the background gas takes place, which typically is ambient air. Therefore, the use of nanosecond TALIF for the quantification of atomic species can be challenging in these complex systems, where the use of faster laser systems and detector in the picosecond or femtosecond temporal range offers a clear advantage~\cite{Schmidt2015,Schmidt2016,Schmidt2017,Frank2005}.

In this paper, TALIF measurements of O and H atom densities in an APPJ (COST-$\upmu$APPJ~\cite{Golda2016}) in a mixture of helium (He) with small amounts of humidity (H$_2$O) using a tunable picosecond (ps) laser system are presented. The enhanced temporal resolution (compared to conventional ns-laser systems) allows us to directly determine the effective collisional-induced quenching rate of the laser-excited states. Therefore, absolute densities of ground state O and H atoms can be determined without knowledge of the collisional dominated ambient environment at atmospheric pressure. 

We compare measured O and H densities with values calculated with a zero-dimensional plasma-chemical kinetics model. The reaction mechanism has been introduced in previous work~\cite{Schroeter2018}. After obtaining good qualitative and quantitative agreement of absolute densities between simulation and experiments, we use the simulation to further investigate the plasma chemical kinetics, such as formation pathways for O and H, as well as the role of oxygen containing impurities on the plasma chemistry. 

\thispagestyle{empty}

\vspace*{0cm}

\section{Experimental setup.}

\subsection{Atmospheric pressure plasma jet.}

The APPJ investigated in this work is similar to the COST-$\upmu$APPJ, which is described in reference~\cite{Golda2016}. The plasma is ignited in a gas channel of $1\times1$~mm$^{2}$ cross section and $30$~mm length, which is confined by two stainless steel electrodes and two quartz windows from the gas inlet to the exit nozzle. One of these electrodes is powered by applying radio-frequency voltage (frequency 13.56~MHz) using a power generator (Coaxial Power Systems RFG-50-13) and an impedance matching unit (Coaxial Power Systems MMN-150-13), while the other is grounded. In this work, we apply a peak-to-peak voltage of 510~V, which is monitored using a high voltage probe (PMK, PPE20KV, 100~MHz). At these low voltages, the plasma operates in $\Omega$~mode \cite{Bischoff2018}. In contrast to the COST-$\upmu$APPJ, the plasma source used in this work does not contain the internal resonance coupler described in \cite{Golda2016} (section 3.2). However, the critical dimensions and operating conditions of both sources are practically the same.

Gas is introduced into the confined discharge channel, and exits at the nozzle into open air. High purity He (99.996\% purity) at a flow of 0.5~slm serves as a buffer gas, and water vapor of up to 0.5\% (5000 ppm) can be admixed in order to create RS due to dissociation of these H$_2$O molecules. The flow rate is chosen to match the conditions of earlier work~\cite{Schroeter2018}, where a ten times higher gas flow was used in a plasma source with an approximately ten times larger cross sectional area compared to the COST-$\upmu$APPJ. For He, the gas flow is regulated by using two mass flow controllers (MFC), as described in earlier work~\cite{Schroeter2018}. One of the two branches is guided through a bubbler, which consists of a 120~cm long domed glass adapter (Biallec GmbH) that is clamped to a KF40 flange with inlet and outlet tubes, as described previously~\cite{Schroeter2018}. The flows of dry and humidified He are combined and fed into the discharge channel. Assuming that the He is saturated with water vapour after passing through the bubbler, the total amount of water in the vapour phase can be calculated using the vapour pressure $p_{\text{H}_2\text{O}}^{\text{vap}}$ of H$_2$O~\cite{Alduchov} and the flow rate of the He through the bubbler $F_{\text{He}}^{\text{bubbler}}$, as has been described elsewhere~\cite{Benedikt2016}:
\begin{align}
p_{\text{H}_2\text{O}}^{\text{vap}} &= 6.112\times 10^{-3}\exp{\left({\frac{17.62T_\text{w}}{243.12+T_\text{w}}}\right)} \text{ bar}\:,\\
F_{\text{H}_2\text{O}} &= F_{\text{He}}^{\text{bubbler}} \frac{p_{\text{H}_2\text{O}}^{\text{vap}}}{p_{\text{atm}}-p_{\text{H}_2\text{O}}^{\text{vap}}}\:,
\end{align}
\noindent where $T_\text{w}$ is the water temperature in $^\circ$C. For some of the measurements carried out here, the bubbler was immersed in a water bath which was regulated at 18$^\circ$C. It will be clarified in each section for which measurements the water cooling has been applied.

\subsection{Picosecond two-photon absorption laser induced fluorescence.}

\subsubsection{Absolute density calibration}

For quantifying the absolute atomic oxygen and hydrogen densities $n_\text{x}$, we measure the spatially, temporally, and spectrally integrated fluorescence signal $S_\text{F}$, and compare its intensity with the fluorescence signal obtained from a noble gas of a known quantity. For the calibration measurement, the plasma source is replaced with a Starna Spectrosil Fluorometer Cuvette, which is filled with the respective noble gas (xenon or krypton) at defined pressures (10~Torr for Xe and 1~Torr for Kr). 

For the further discussion, the following abbreviations are used when mentioning different states: ``O" for the ground states $\sum_J$O(2p$^4$ $^3$P$_J$) and ``H" for the H(1s $^2$S$_{1/2}$) ground state. The excited states are abbreviated as O$^*$ for the $\sum_J$O(3p $^3$P$_J$) (or short: O(3p $^3$P)), and as H$^*$ for the H(n=3) state.

A comparison of the fluorescence intensities yields absolute densities for the species of interest
\begin{equation}
\frac{S_\text{F,x}}{S_\text{F,cal}} = \frac{\eta(\lambda_\text{F,x})}{\eta(\lambda_\text{F,cal})} 
\frac{T_\text{f}(\lambda_\text{F,x})}{T_\text{f}(\lambda_\text{F,cal})}
\frac{1}{T_\text{c}(\lambda_\text{F,cal})}
\frac{a_{ik\text{,x}}}{a_{ik\text{,cal}}}
\frac{\sigma_\text{x}^{(2)}}{\sigma_\text{cal}^{(2)}}
\frac{n_\text{x}}{n_\text{cal}}
\left(\frac{E_\text{x}}{E_\text{cal}/T_\text{c}(\lambda_\text{L,cal})}
\frac{\lambda_\text{L,x}}{\lambda_\text{L,cal}}\right)^2 \:.\label{eq:TALIF_cal} \\
\end{equation}
Here, $\lambda$ is the wavelength of either the laser (L) or fluorescence (F) radiation for the species of interest (x = O, H) and calibration species (cal = Xe, Kr), $T_\text{c}$ is the transmission of the calibration cuvette which contains the noble gas, $T_\text{f}$ the transmission of the interference filter placed in front of the camera for the respective wavelength, $\sigma^{(2)}$ the two-photon excitation cross section, and $E$ the laser pulse energy. If the laser-excited, fluorescent state is denoted by the letter $i$, then $a_{ik}$ denotes the branching ratio of the transition into the lower state $k$
\begin{align}
a_{ik}&=\frac{A_{ik}}{\sum_k A_{ik} + \sum_q k_q^i n_q} \notag\\
&= \frac{A_{ik}}{A_i + Q_i} \notag\\
&= \frac{A_{ik}}{A_i} \frac{A_i}{A_i+Q_i} \notag\\
&= b_{ik} \frac{\tau_\text{eff}}{\tau_\text{nat}}\:.
\end{align}
$A_{ik}$ is the decay rate for the transition from state $i$ to $k$, while the inverse of the sum of decay rates into all possible lower states $k$ is the natural lifetime $\tau_\text{nat}=(\sum_k A_{ik})^{-1}$ of the state $i$. The natural lifetimes for O$^*$, H$^*$, Xe(6p'[3/2]$_2$), and Kr(5p'[3/2]$_2$) are 34.7~ns, 17.6~ns, 40.8~ns, and 34.1~ns, respectively, according to~\cite{Niemi2001,Niemi2005}. $b_{ik}$ is the purely optical branching ratio into a specific state $k$, and $\tau_\text{eff}$ the effective lifetime of the laser excited state. The effective lifetime takes into account both the natural lifetime of the excited state, as well as radiation-less collisional de-excitation (quenching) via collisions of the excited states with the background gas, which can significantly lower the excited state lifetime. Quenching is dependent on quenching coefficients $k_q$ and densities $n_q$ of the quenching species.

Comprising all instrumental constants into one overall calibration constant $C$, \cref{eq:TALIF_cal} can be further simplified to
\begin{equation}
n_\text{x}=C
\frac{\tau_\text{eff,cal}}{\tau_\text{eff,x}}
\frac{S_\text{F,x}}{S_\text{F,cal}} 
n_\text{cal}\:, \label{eq:TALIF_cal_short}
\end{equation}
where $S_\text{F}$ and $\tau_\text{eff}$ are the measurable quantities. 

\begin{figure}[ht]  
	\centering
	.pdf
	\includegraphics[width=13cm]{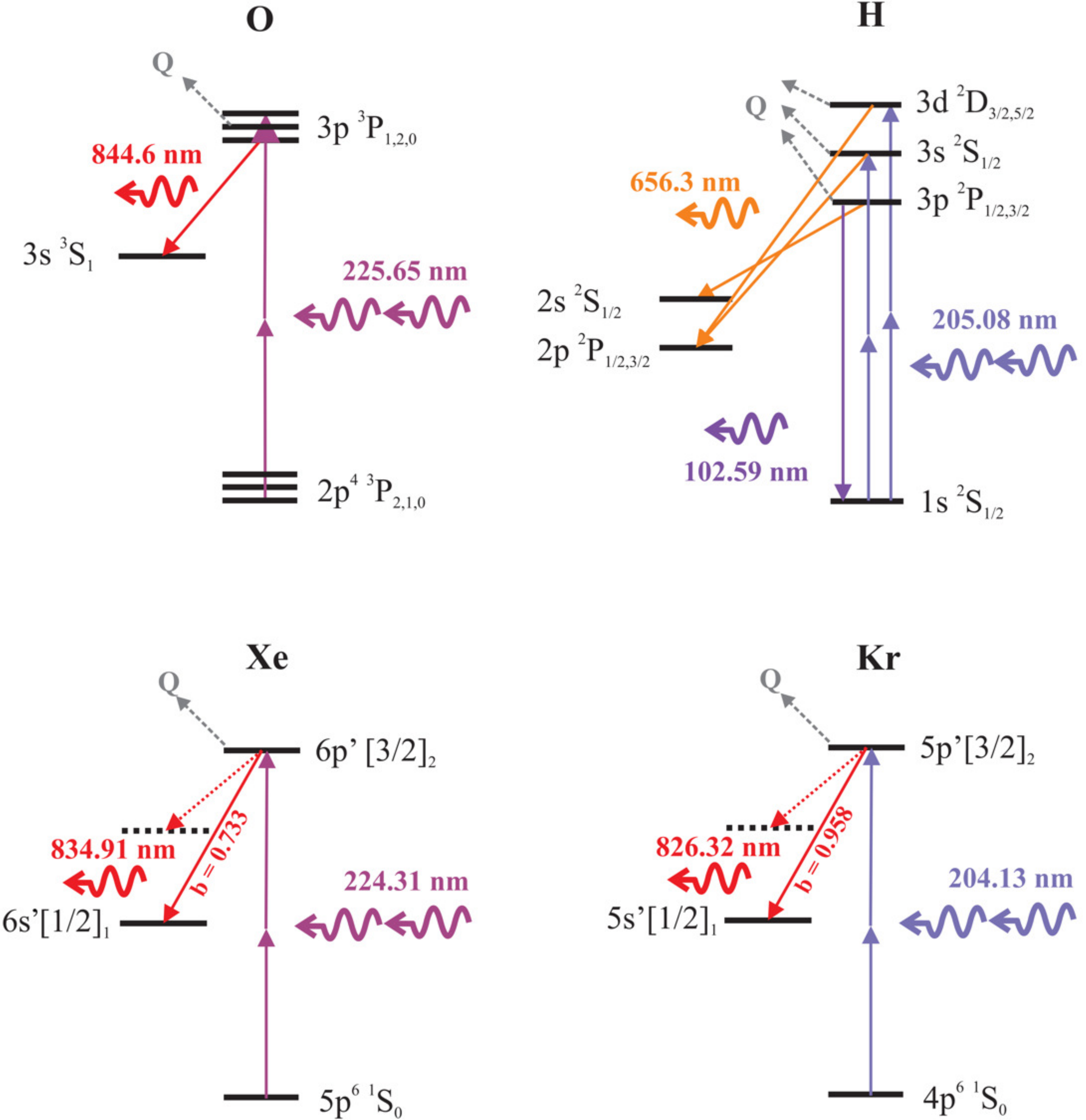}
	
	\caption[TALIF energy schemes for O, N, Xe, and Kr]{TALIF schemes for the investigated species O and H and corresponding calibration gases, which are Xe and Kr, respectively.}
	\label{fig:LIF_schematics}
\end{figure}

The established calibration schemes for O and H using Xe and Kr~\cite{Niemi2001,Niemi2005} are shown in \cref{fig:LIF_schematics} including the emission and fluorescence wavelengths, and purely optical branching ratios $b_{ik}$. Data for the schematics have been taken from other publications~\cite{Niemi2001,Niemi2005} and the NIST Atomic Spectra Database~\cite{NIST_atom}.

The calibration schemes are chosen in a way such that both the excitation and fluorescence wavelengths for the species of interest and calibration species are close, to prevent any influence of changes in transmission or beam profile due to optics used to guide the laser beam. The schematics presented in \cref{fig:LIF_schematics} have been discussed in much detail in previous work~\cite{Boogaarts2002,Reuter2009,Gessel2013,Schmidt2016}, and only the main aspects will be discussed here.

In \textit{atomic oxygen}, both the ground and the TALIF excited state (O(2p$^4$ $^3$P$_J$) and O(3p $^3$P$_J$), respectively) are split into three energy sub-levels $J=0,1,2$. While the ground state levels have a distinct energy gap in the order of a few hundred wavenumbers, the upper states lie energetically very close, i.e. within one wavenumber. Therefore, excitation from one of the ground state levels can populate all three $J=0,1,2$ upper state levels, if optically allowed, because their splitting is within the spectral width of the laser used in this work (about 4~cm$^{-1}$). 

\textit{Xenon} is typically chosen as a calibration gas for O, since both the excitation and fluorescence wavelengths are in close proximity. The excited Xe state can decay into several sublevels. $b_{ik}$ for the transition into the 6s state with a wavelength of 834.9~nm is 0.733~\cite{Horiguchi1981}. 

In this work, the following two-photon absorption cross section ratio is used for O and Xe~\cite{Niemi2005}
\begin{equation}
	\frac{\sigma_\text{Xe}^{(2)}}{\sum_{J'}\sigma_{\text{O},J\rightarrow J'}^{(2)}} = 1.9 (\pm 20\%) \:,
\end{equation}
which takes into account all transitions into the upper excited states of atomic oxygen. In the measurements presented in this work, usually only the lowest $J=2$ sub-level of the electronic ground state of O is probed. Particularly for measurements where humidity is added to the gas flow, the fluorescence signal from the other sub-levels is too weak to be detected reliably. In order to calculate the absolute density of all ground state sub-levels, the theoretical Boltzmann factor is applied
\begin{align}
	\frac{n_2}{n_\text{O}}=\frac{g_2}{\sum_{J=0..2} g_J \exp{(-E_J/k_B T_g)}}\:. \label{eq:Boltz}
\end{align}
Here, $n_2$ is the density of the probed sub-level of the ground state, and n$_\text{O}$ is the density of the sum of all three ground levels. $g$ is the statistical weighting, $E_J$ the energy of the respective sub-level of the ground state, and $T_g$ the gas temperature. The latter was measured with a thermocouple, about 315~K under our experimental conditions.

In \textit{atomic hydrogen} the 3s and 3d sub-levels can be excited by the linearly polarised laser radiation at 205.08~nm according to the selection rules for two-photon absorption transitions, but not the 3p sub-level. The spectral separation between the 3s and 3d states is about 0.15~cm$^{-1}$, so well within the laser bandwidth of about 4 cm$^{-1}$. The natural lifetimes of the 3s and 3d states are 159~ns and 15.6~ns, respectively~\cite{Condon1951,Preppernau1995}, while the theoretical ratio of the excitation cross sections $\frac{\sigma_d^{(2)}}{\sigma_s^{(2)}}$ is 7.56~\cite{Preppernau1995}. Important for the TALIF calibration is that we use the natural lifetime $\tau_\text{nat}=17.6$~ns, as measured in~\cite{Niemi2001}, resulting from the weighted combination of the 3s and 3d states.

\textit{Krypton} is the gas that is typically used to calibrate TALIF measurements of atomic hydrogen. Similar to the Xe-O calibration, the excitation wavelengths of Kr and H are spectrally close, as shown in \cref{fig:LIF_schematics}. The fluorescence wavelengths differ by almost 200~nm, which results only in an insignificant change of the focal length of approximately 50~$\upmu$m for the particular lens used in the ps-TALIF setup. The setup will be described in the next section.

For the two-photon excitation cross section ratio, the following value is used~\cite{Niemi2001}
\begin{equation}
	\frac{\sigma_\text{Kr}^{(2)}}{\sigma_\text{H}^{(2)}} = 0.62 (\pm 50\%)\:. \\
\end{equation}
Similar to Xe, the cuvette is filled with Kr for the calibration measurement. The pressure in the calibration cell is chosen as 1~Torr. At higher pressures, there is the risk that Amplified Spontaneous Emission (ASE) disturbs the fluorescence characteristic and decay, as previously observed~\cite{West2016}. The purely optical branching ratio for the Kr transition of interest is $b=0.953$~\cite{Chang1980}.

In the following paragraphs the measurement protocols for theses quantities, as well as possible saturation effects, are discussed in detail.

\subsubsection{Experimental apparatus}

\begin{figure*}[]
	\begin{center} 
		\includegraphics[width=0.8\textwidth]{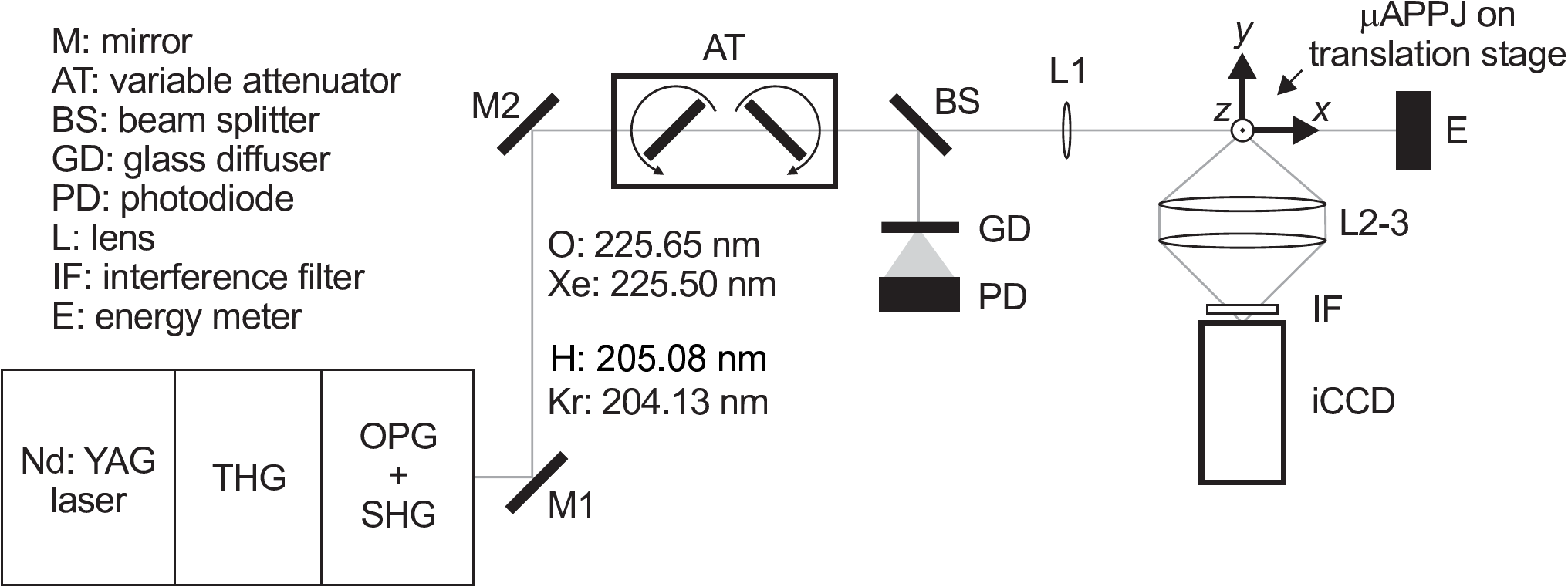}
		\hspace{0.4cm}
	\end{center}
	\caption{Schematic of the experimental setup depicting the picosecond laser system and the components used for beam energy control and fluorescence detection. The APPJ is mounted on a translation stage for spatial mapping of the effluent in the y-z plane.}
	\label{fig:TALIF_experiment} 
\end{figure*}

The modular laser set-up (EKSPLA) is shown in fig.~\ref{fig:TALIF_experiment}. It includes a Nd:YAG pump laser that incorporates a mode-locked oscillator together with regenerative and power amplifiers. The beam is directed into an amplification and harmonics generation unit, and subsequently enters an optical parametric generator and amplifier followed by sum-frequency and difference-frequency generation, which offers spectral tunability within the range from 193 to 2300~nm.
In the UV range, the laser system generates pulses of 30~ps duration, a few hundreds of~$\upmu$J pulse energy, and a spectral width of approximately four wave numbers. The laser pulse energy is varied with help of an attenuator-compensator system, which comprises two coated counter-rotating CaF$_{2}$ substrates that are controlled by a stepper motor. The standard deviation of the shot-to-shot fluctuations in the pulse energy is about 8\%.

The laser beam is focused by a spherical plano-convex fused-silica lens with 30~cm focal length in a plane intentionally chosen about a centimetre behind the plasma effluent. This helps spreading the spatial laser pulse energy/power over a larger volume, resulting in a less stringent saturation level of the two-photon transition at the cost of a lower overall fluorescence signal, as well as staying below the material damage threshold of the calibration cuvette. The fluorescence radiation of the excited states is detected in the direction perpendicular to the laser beam using an intensified charge coupled device camera (iCCD: Stanford Computer Optics 4-Picos, $780 \times 580$ array, 8.3~$\upmu$m$^{2}$~pixels, $\text{S}25\text{IR}$ photo-cathode) subsequent to its passage through a doublet of achromatic lenses~(diameter 50~mm, focal length 80~mm each) and an interference filter~(central wavelengths~$\lambda_{\text{O}}=845$~nm, $\lambda_{\text{H}}=656$~nm, $\lambda_{\text{Xe}}=835$~nm, $\lambda_{\text {Kr}}=825$~nm, each with a full-width-at-half-maximum of 10~nm. 

\subsubsection{Signal measurement.}

\begin{figure}[h]  
	\begin{subfigure}[h]{0.48\linewidth}
		\includegraphics[width=\textwidth]{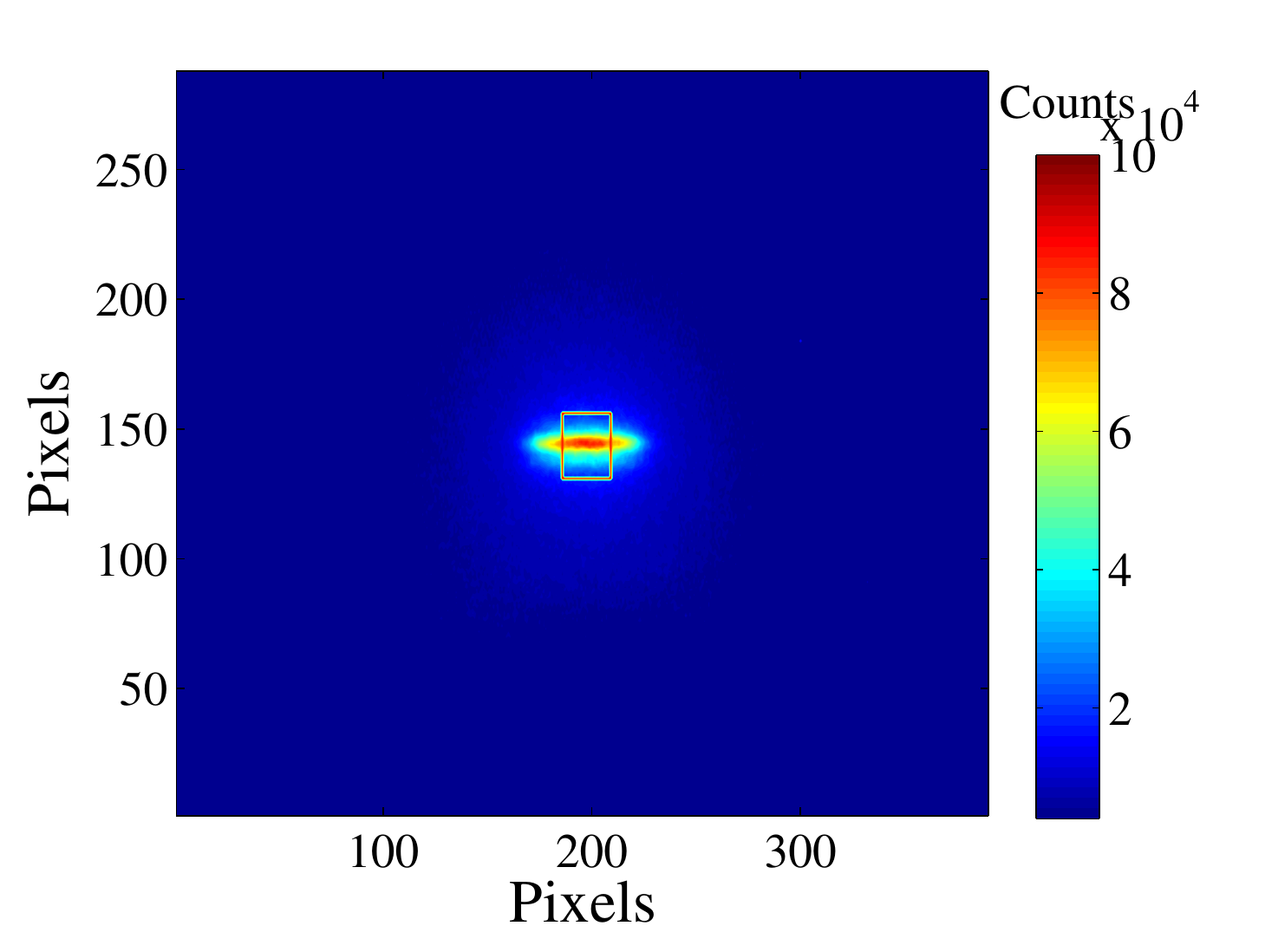}\llap{\parbox[b]{3in}{(a)\\\rule{0ex}{1.8in}}}	
	\end{subfigure}
	\hfill
	\begin{subfigure}[h]{0.48\linewidth}
		\includegraphics[width=\textwidth]{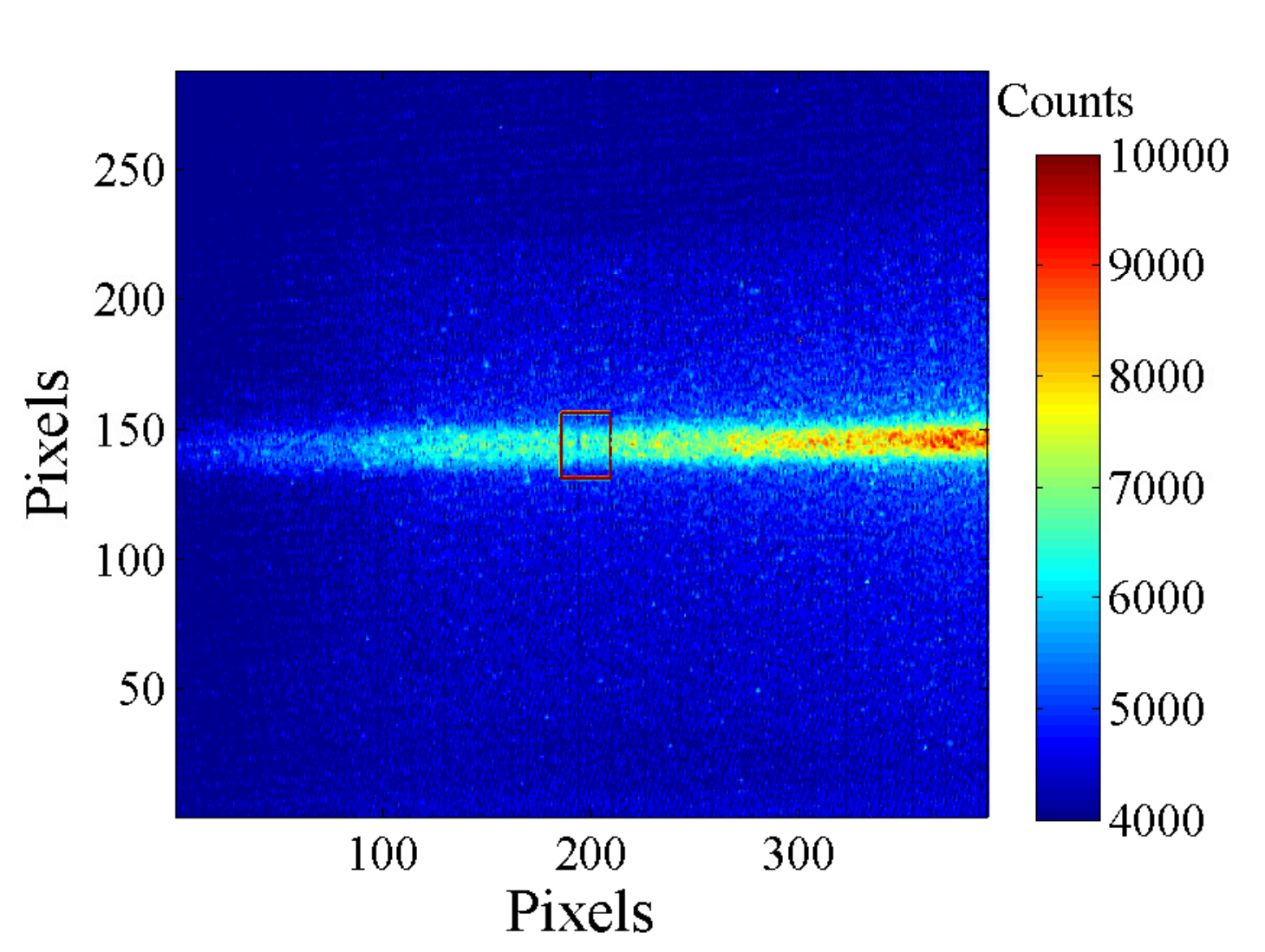}\llap{\parbox[b]{3in}{(b)\\\rule{0ex}{1.8in}}}
	\end{subfigure}
	\caption[Fluorescence images for O and Xe, and position of $\upmu$APPJ]{Example of fluorescence images taken with the ICCD camera: (a) for atomic oxygen, (b) for xenon. The red rectangular shape indicates the ROI.}
	\label{fig:TALIF_signal}
\end{figure}

As already mentioned, $S$ is the spatially, temporally, and spectrally integrated fluorescence signal. The spatial integration is performed by choosing a defined region of interest (ROI) from the camera image, in which the signal is summed up. This means that the spatial resolution is limited by the choice of ROI. \Cref{fig:TALIF_signal}~(a) shows the fluorescence signal of O$^*$ at 844~nm, at the strongest excitation wavelength of 225.65~nm. A bright region is visible approximately in the middle of the camera image where the laser beam intersects with the jet effluent region, leading to excitation of O from the ground state, and subsequent emission of fluorescence. Similarly, \cref{fig:TALIF_signal}~(b) shows the fluorescence signal obtained from the calibration with Xe at 835~nm and a laser wavelength of 224.30~nm. The calibration cuvette containing the Xe gas has much larger dimensions compared to the 1~mm discharge gap of the $\upmu$APPJ, therefore, the fluorescence signal is visible over the whole width of the CCD chip. For both measurements, the same ROI is chosen, as indicated in the figures.

The temporal integration is performed by choosing the camera gate width long enough to collect almost all of the exponential fluorescence decay after the laser pulse ($>$98\% under all measurement conditions). Among the four species of interest, we measured the longest effective lifetime of 21.4~ns for Kr(5p'[3/2]$_2$) at 1~Torr in the cuvette. Therefore, a camera gate width of 100~ns is chosen, which means that 98-99\% of the fluorescence signal is captured, depending on the camera start delay. The same gate width is chosen for all other species, leading to a light capture higher than 99.9\%.  

\begin{figure}[ht]  
	\centering
	
	\begin{subfigure}[h]{0.47\linewidth}
		\centering
		\includegraphics[width=\textwidth]{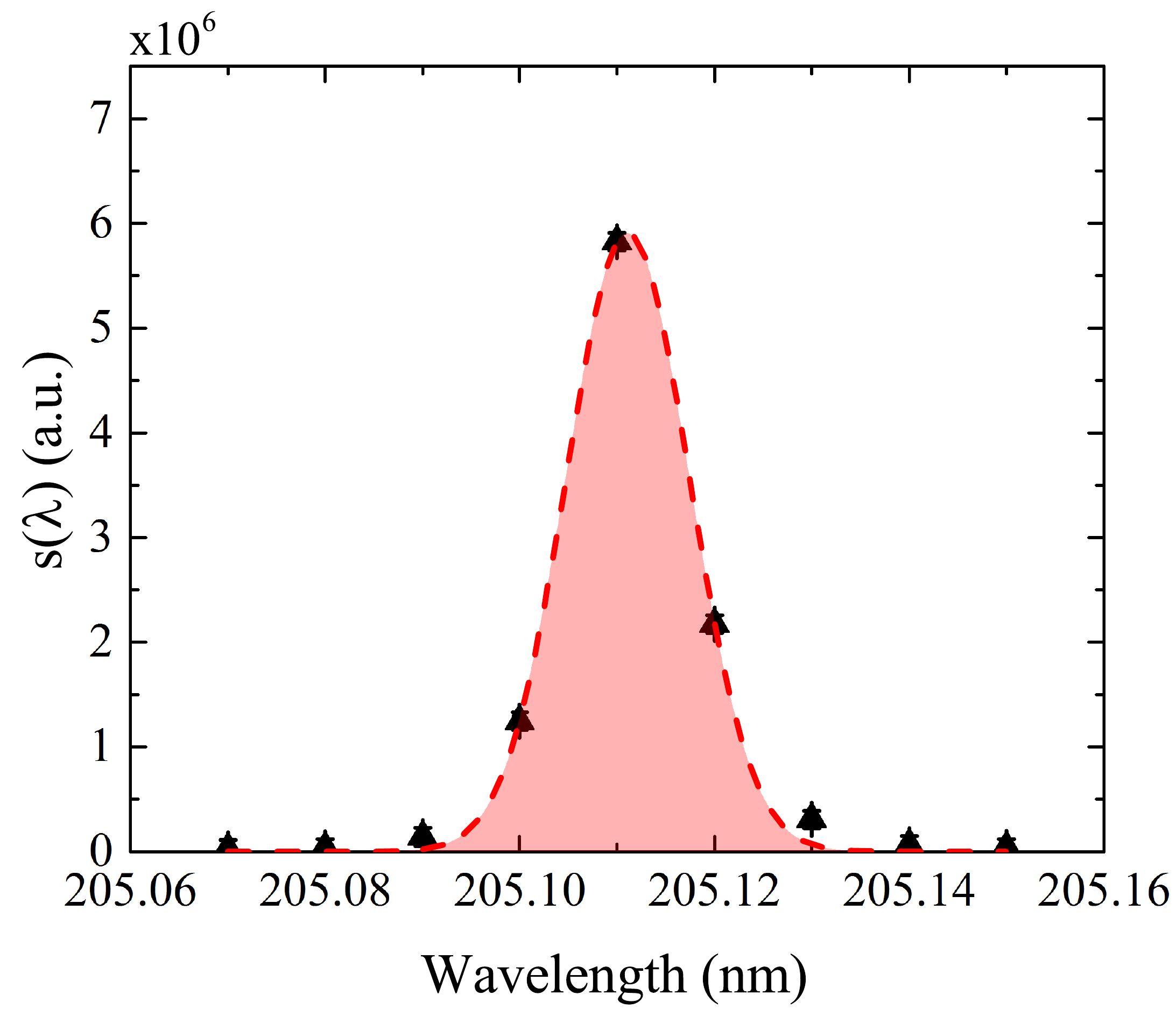}\llap{
			\parbox[b]{2.8in}{(a)\\\rule{0ex}{2in}
			}} 
	\end{subfigure}
	\hfill
	\begin{subfigure}[h]{0.47\linewidth}
		\centering
		\includegraphics[width=\textwidth]{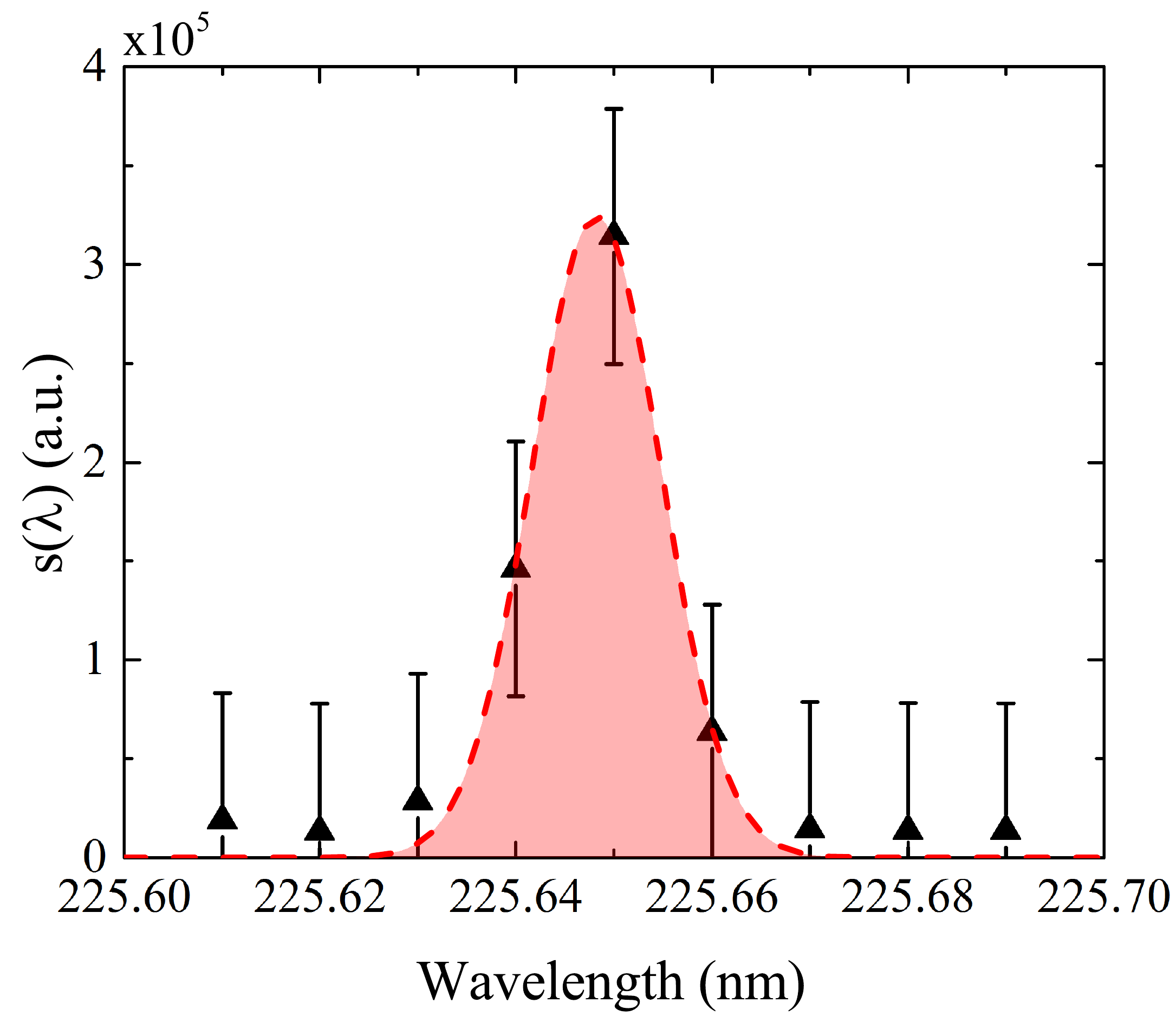}\llap{
			\parbox[b]{2.8in}{(b)\\\rule{0ex}{2in}
			}}
	\end{subfigure}
	\caption[Fluorescence signal for O(3p $^3$P) with O$_2$ and H$_2$O admixtures]{Normalised fluorescence signal for (a) H$^*$ and (b) O$^*$ 500~sccm He with a H$_2$O content of 1240~ppm, 510~V$_\text{pp}$. Error bars show the standard deviation of the noise. The data is fitted with a Gaussian function.}
	\label{fig:TALIF_scan}
\end{figure}

Temporal and spatial integration of the fluorescence signal yields the spectrally dependent fluorescence signal $s(\lambda)$, which is a function of laser wavelength over the resonant transition. The minimum wavelength step of the laser is 0.01~nm when tuned. With this spectral resolution we measure at 9 different laser wavelengths around the absorption line. Additionally, a measurement of the background signal is performed by manually closing the shutter of the laser output. By subtracting this background signal, potential background light from the plasma source is accounted for, as well as base noise from the detector. Spectral integration of $s(\lambda)$ results in the absolute fluorescence signal $S_\text{F}$. Typical wavelength scans for atomic hydrogen and oxygen are shown in \cref{fig:TALIF_scan}.

The overall fluorescence signal is determined as the area of a Gaussian function that is fitted to the measured line profile $s(\lambda)$. The overall line profile is dominated by the laser line profile, whose bandwidth as stated by the manufacturer is about 4~cm$^{-1}$ ($\Delta\lambda=20$~pm). This is less than ten times the theoretical Fourier limit, when considering the pulse duration of 32~ps. The other broadening mechanisms, i.e. Doppler and pressure broadening, are about an order of magnitude smaller under our experimental conditions. 

For atomic oxygen, the Doppler broadening $\Delta\lambda_\text{D}$ can be calculated as
\begin{equation}
	\Delta \lambda_\text{D}^\text{O} = \frac{\lambda}{c}\sqrt{\frac{8k_BT_g \ln{(2)}}{m_O}} = 0.35 \text{ pm}\:,
\end{equation}
where $T_\text{g}=315$~K and $\lambda=112.8$~nm as half the excitation wavelength for atomic oxygen, which is in good agreement with measured values~\cite{Marinov2016}. For atomic hydrogen, $\Delta\lambda_\text{D}^\text{H}=1.3$~pm, which is still more than a factor 10 smaller than the laser bandwidth.
		
Pressure broadening coefficients for O by He and O$_2$ were determined in reference~\cite{Marinov2016}. For 1 bar, pressure broadening by the background He can be calculated as
\begin{equation}
	\Delta\lambda_\text{L}^\text{O} = 0.59 \text{ pm}\:,
\end{equation}
assuming that pressure broadening is dominated by the He background gas.

In general, our pico-second laser system offers a balanced compromise between temporal and spectral resolution when compared to typical nano-second and femto-second UV TALIF laser systems~\cite{Schmidt2017}.

The quality of the signal measurement strongly depends on the investigated species and the experimental circumstances, and the signal-to noise ratio, which is defined here as
\begin{align}
	SNR = \frac{s-s_0}{\Delta noise}\:,
\end{align}
where $\Delta noise$ is the standard deviation of the measured noise, and $s_\text{net} = s-s_0$ the net mean fluorescence signal $s$, which means the signal averaged over the ROI minus the average background $s_0$. For example, when measuring atomic hydrogen under a H$_2$O variation, the measured H fluorescence signal is strong since H densities produced from H$_2$O are high. \Cref{fig:TALIF_scan}~(a) shows normalised fluorescence signal of H$^*$ as a function of the laser wavelength. The SNR for this measurement is good, resulting in a small $\Delta noise$ (shown as error bars) compared to the signal strength. On the contrary, for a measurement of O under a H$_2$O admixture, the signal to noise ratio is small, because the O densities produced from H$_2$O are low. This is shown in \cref{fig:TALIF_scan}~(b). 
		
As discussed previously, the laser steps are limited to 0.01~nm, resulting in typically 4-5 possible measurements where a signal is obtained for each wavelength scan. Although a Gaussian function can usually be fitted to the experimental data with no difficulties, the fact that only so few points are available for the fit makes it difficult to assess the accuracy of the fitting procedure, particularly when the SNR is low.

\subsubsection{Lifetime measurement.}

In order to measure $\tau_\text{eff}$ in \cref{eq:TALIF_cal_short} with the ps-TALIF setup, the gate width of the camera is fixed, and the camera delay is increased incrementally, so that the fluorescence signal $s(\lambda)$ at different times after the laser pulse is obtained.

\begin{figure}[ht]  
	\begin{subfigure}[h]{0.45\linewidth}
		\centering
		\includegraphics[width=\textwidth]{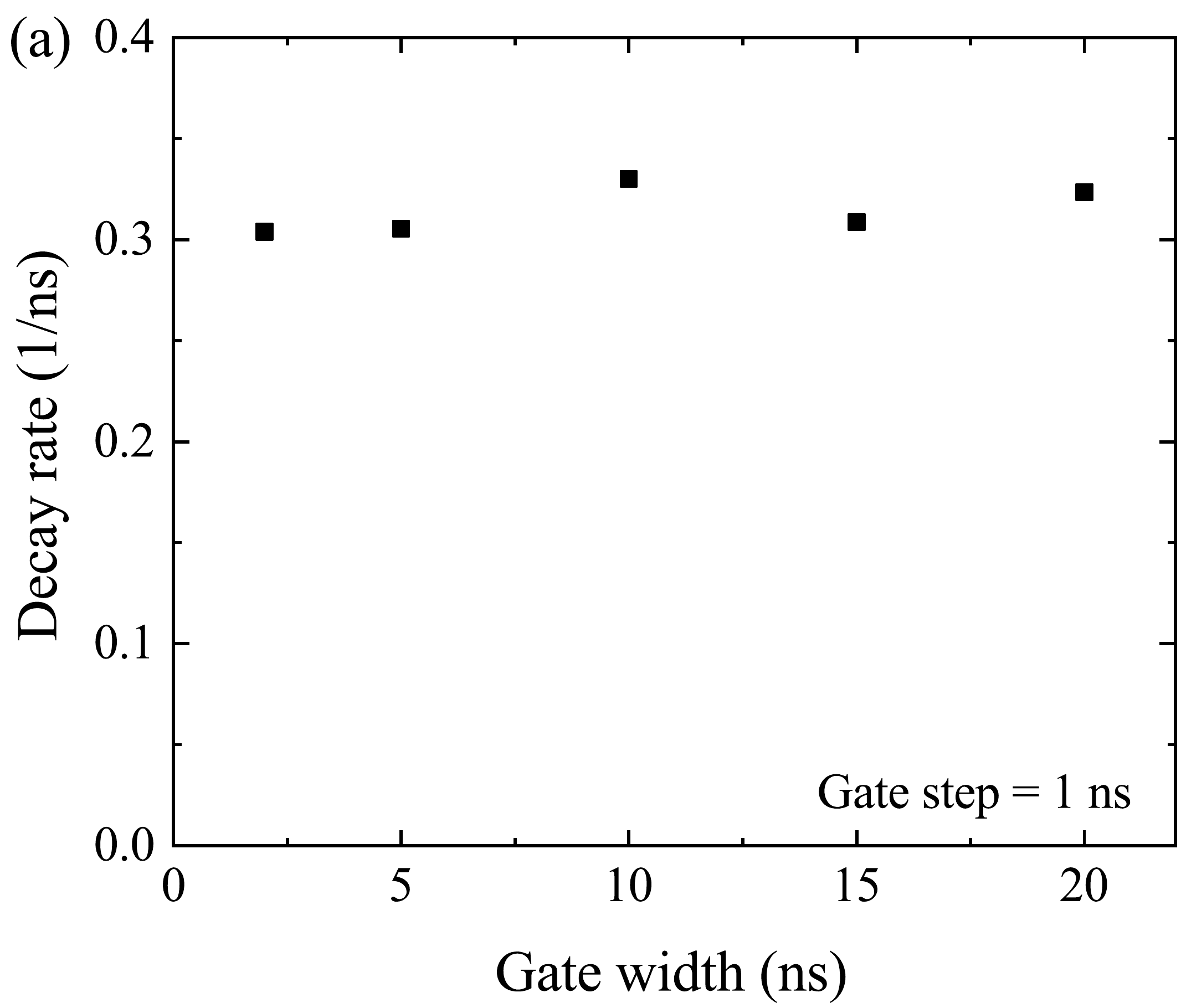}
	\end{subfigure}
	\hfill
	\begin{subfigure}[h]{0.46\linewidth}
		\centering
		\includegraphics[width=\textwidth]{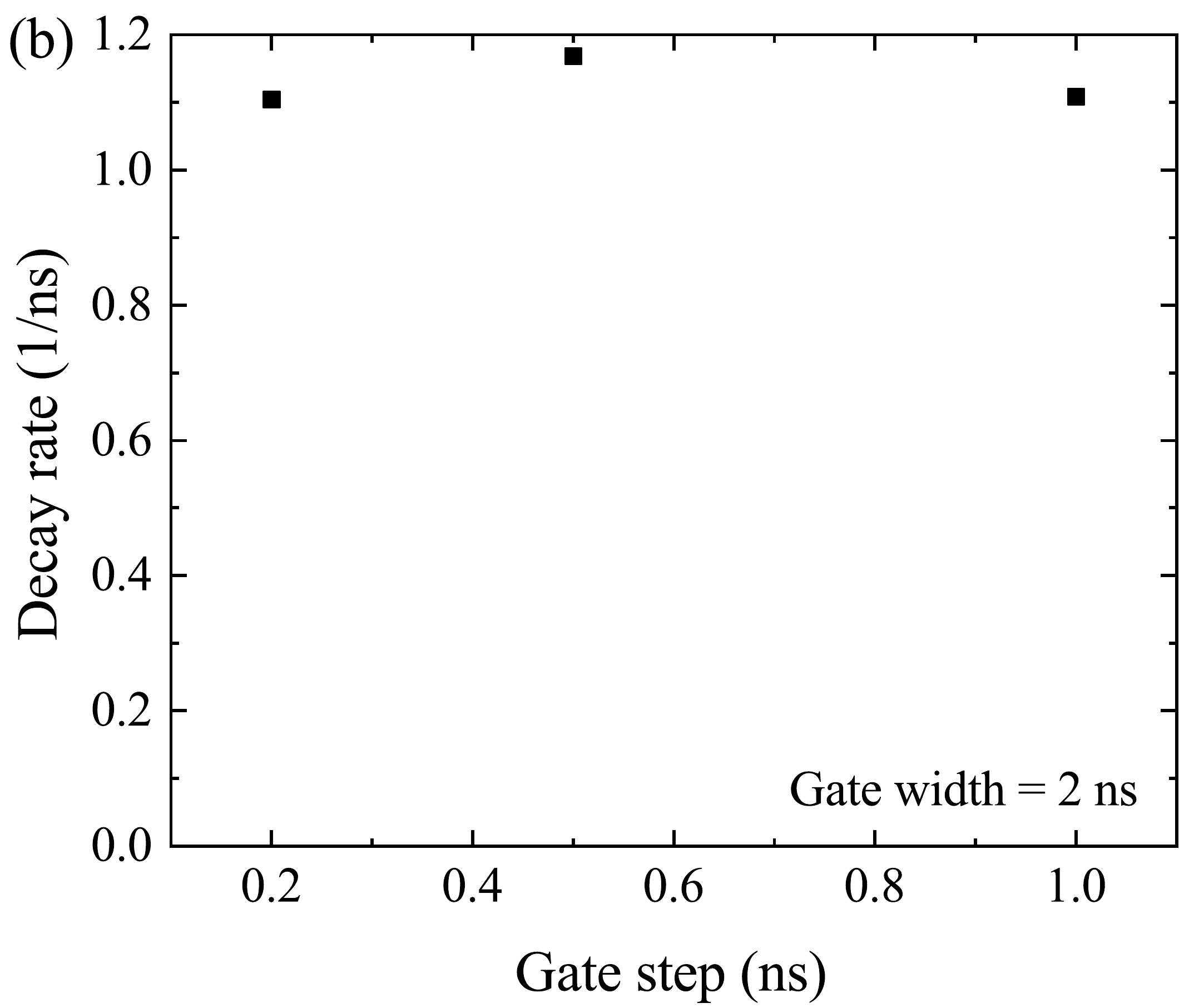}
	\end{subfigure}
	
	\caption{Measured decay rates of (a) O$^*$ as a function of gate width at constant gate step (0.5~slm He, 4130~ppm H$_2$O) and (b) H$^*$ as a function of gate step at a constant gate width (1~slm He, 1040~ppm H$_2$O).}
	
	\label{fig:gate}
	
\end{figure}

Typical camera gate steps are chosen as 0.5~ns for O$^*$, 0.2~ns for H$^*$, 1~ns for Xe(6p'[3/2]$_2$), and 2~ns for Kr(5p'[3/2]$_2$). Typical gate widths are chosen as 10~ns for O$^*$, 2~ns for H$^*$ and Xe(6p'[3/2]$_2$), and 10~ns for Kr(5p'[3/2]$_2$), respectively. Since the camera gate widths are larger than the respective gate steps, the measured signals overlap temporally. Particularly the long gate widths for the lifetime measurements of  O$^*$ and Kr(5p'[3/2]$_2$) have been chosen due to the low signal-to-noise ratio for these measurements. It was tested before for the same plasma operating conditions, that the choice of gate step (between 0.2 and 1 ns) and width (between 2 and 20 ns) does not have an effect on the measured lifetimes (see \cref{fig:gate}). 

%
%
%

\begin{figure}[ht]  
 	\begin{subfigure}[h]{0.45\linewidth}
 		\centering
 		\includegraphics[width=\textwidth]{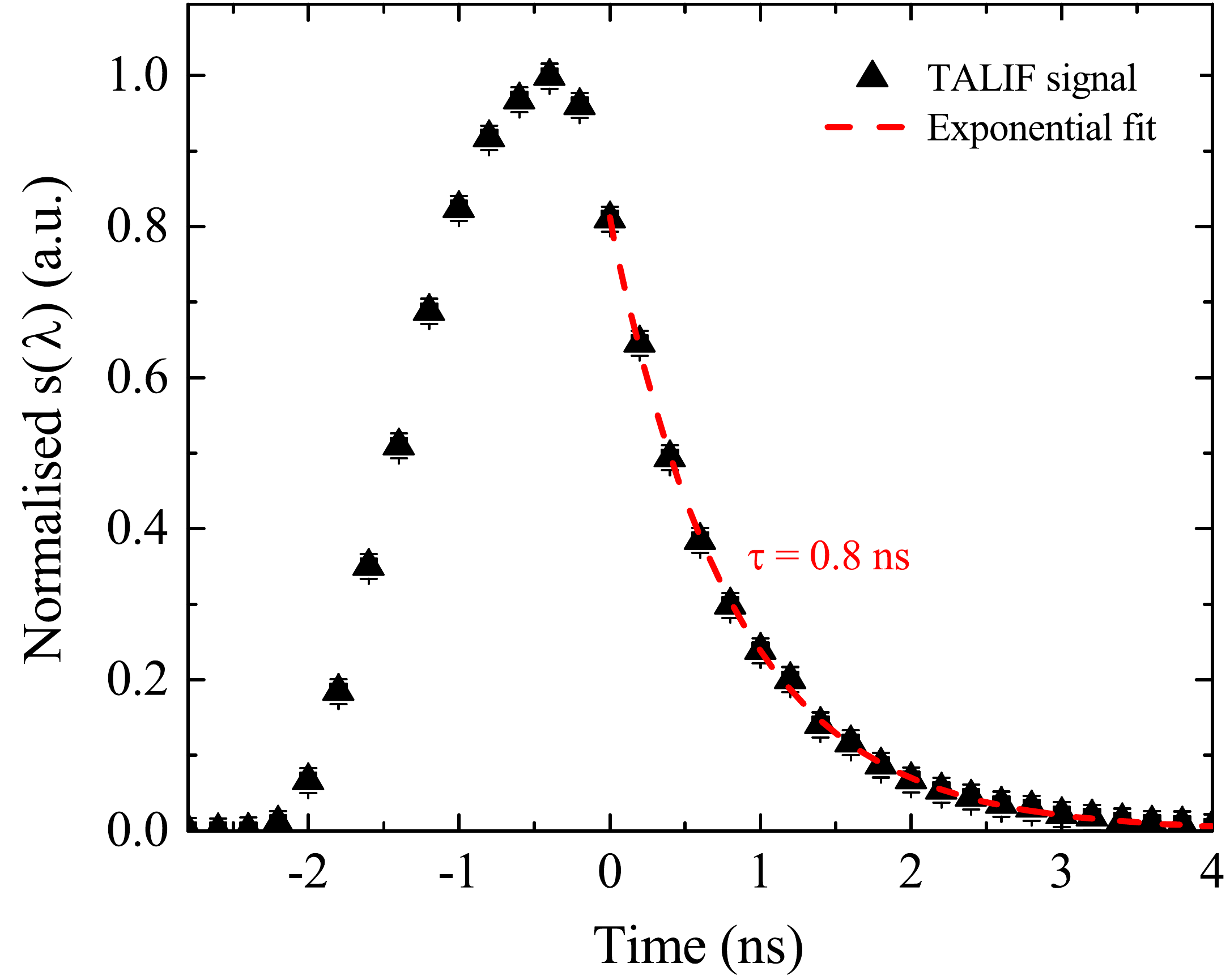}\llap{
 			\parbox[b]{2.2in}{(a)\\\rule{0ex}{1.95in}
 			}} 
 		\end{subfigure}
 		\hfill
 		\begin{subfigure}[h]{0.46\linewidth}
 			\centering
 			\includegraphics[width=\textwidth]{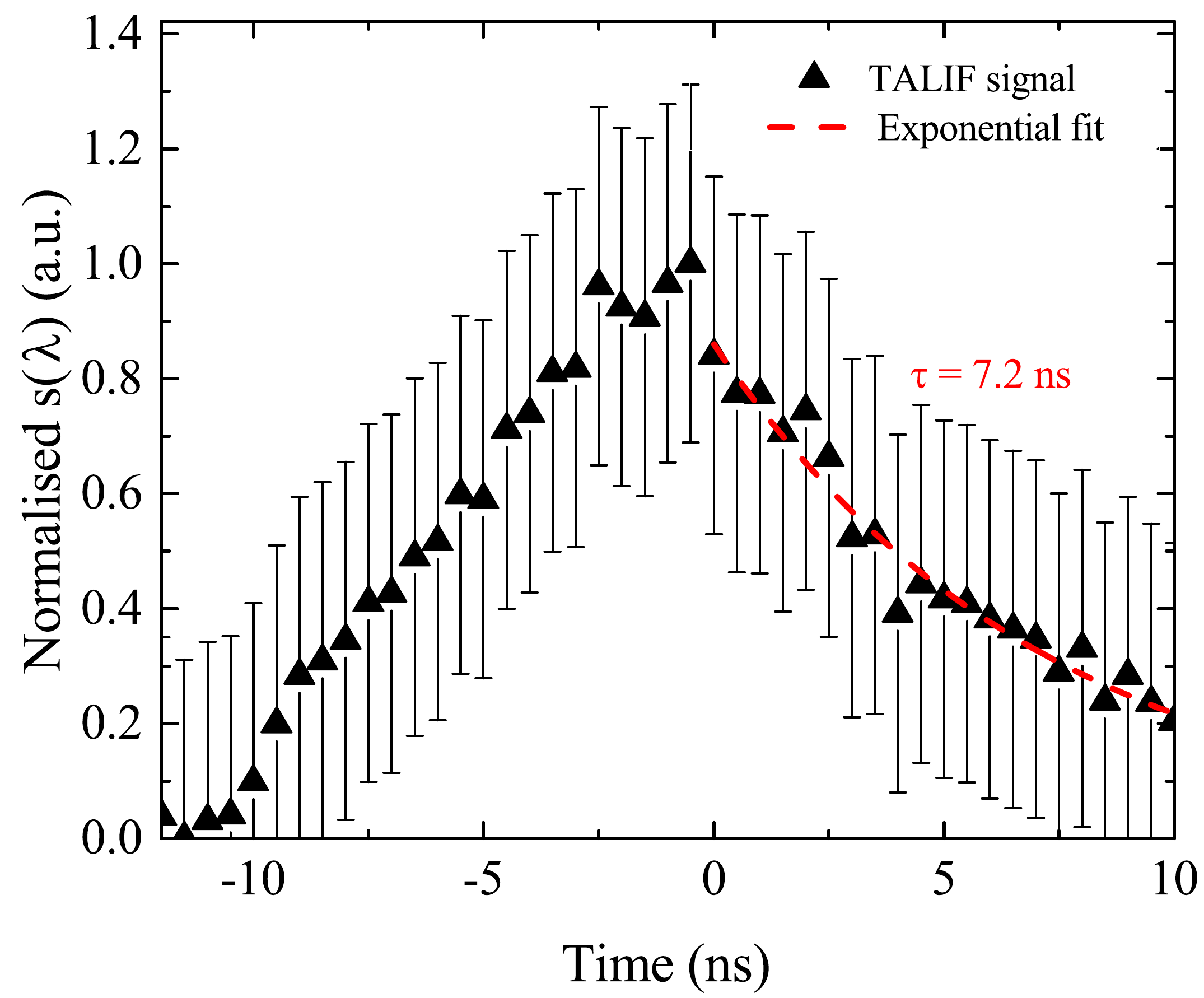}\llap{
 				\parbox[b]{2.3in}{(b)\\\rule{0ex}{1.95in}
 				}}
 			\end{subfigure}
 			
 			\caption[Normalised measured fluorescence decay for O(3p $^3$P) with O$_2$ and H$_2$O admixtures]{Normalised measured fluorescence decay for (a) H$^*$ and (b) O$^*$ at 500~sccm He, H$_2$O content of 1240~ppm, 510~V$_\text{pp}$. Error bars show the standard deviation of the noise.}
 			
 			\label{fig:SNR_quench}
 			
 		\end{figure}

As for the signal measurements, the quality of the lifetime measurement strongly depends on the investigated species and the experimental circumstances, and the signal-to-noise ratio. The lifetime of the excited species can be obtained by fitting an exponential decay to the measured TALIF signal. This is shown in \cref{fig:SNR_quench} for measurements of the atomic hydrogen and oxygen fluorescence for the same plasma conditions. Similar to the previous discussion, atomic hydrogen is produced in large quantities from H$_2$O (\cref{fig:SNR_quench}~(a)), resulting in a good signal-to-noise ratio, whereas O densities produced from H$_2$O are much smaller (\cref{fig:SNR_quench}~(b)), and $\Delta noise$ is large compared to the fluorescence signal. However, a large number of points is generally taken, and the fluctuation of the measurement is small. The decay can be easily fitted by an exponential decay, with stated error bars according to a 95\% confidence band. The effective lifetimes for Xe(6p'[3/2]$_2$) and Kr(5p'[3/2]$_2$) (at 10~Torr and 1~Torr pressure) have been measured each time an absolute calibration was performed. The lifetimes and their uncertainties are therefore calculated as the average and standard deviations from multiple measurements as $\tau_\text{eff}^{\text{Xe(6p'[3/2]}_2\text{)}}=6.5\pm0.5$~ns and $\tau_\text{eff}^{\text{Kr(5p'[3/2]}_2\text{)}}=21.4\pm0.8$~ns, respectively.
 
\subsubsection{Choice of laser energy}

The calibration according to \cref{eq:TALIF_cal} only holds when the observed TALIF signal depends on the square of the laser pulse energy. This is the case for weak enough laser energy to only excite a small amount of the ground state atoms, and not disturb the system otherwise. In reality, several effects can occur, if the laser energy is chosen to be too high, such as photo-dissociation of molecules being present in the gas, or photo-ionisation when a third photon is absorbed by the already excited atom. These effects would lead to a deviation of the square dependence of the TALIF signal and the laser pulse energy. This can be easily checked by varying the laser pulse energy, and plotting the spectrally integrated TALIF signal $S_\text{F}$ against the squared laser pulse energy.

In this work, we can observe saturation effects at higher laser pulse energies for O, Xe, and Kr, while for H no saturation was observed for energies up to 40~$\upmu$J. For all measured species, we choose laser pulse energies that are well below the saturation limit, respectively: 24~$\upmu$J (O), 35~$\upmu$J (H), 0.45~$\upmu$J (Xe), and 0.28~$\upmu$J (Kr). The pulse energies for the probing species (O and H) are much lower compared to ns-TALIF setups, where typical laser energies lie in the range of millijoules~\cite{Niemi2005,Wagenaars2012}. Therefore, comparing the ps-TALIF setup with a standard ns-TALIF setup, average powers are lower for the ps-TALIF setup. Pulse peak powers of the two systems are about in the same order of magnitude ($\sim10^5$~W).

\subsubsection{Constants and error estimation}
				
\Cref{tab:TALIF_constants} shows values used together with \cref{eq:TALIF_cal} to calculate absolute densities of O and H. Values for the natural lifetimes as well as branching ratios have been taken from references~\cite{Niemi2001,Niemi2005}. As discussed previously, the laser energy is usually monitored using an energy meter. Values for the quantum efficiency of the detector were taken as stated by the manufacturer. The spectral transmission of the optical filters were previously measured using a Shimadzu UV-1800 UV-VIS spectrophotometer with 0.1~nm resolution~\cite{West2016}. The transmission of the calibration cuvette was determined by measuring the laser pulse energy in front and behind the cuvette. 
				
\begin{table}[h]
	\centering
	\small
	\caption[Constants used for TALIF calibration]{Constants used for calculation of absolute densities according to \cref{eq:TALIF_cal} for the excitation wavelength $\lambda_\text{L}$, the natural lifetime $\tau_\text{nat}$, the purely optical branching ratio $b$, the laser energy $E_\text{L}$, the quantum efficiency $\eta$, and transmissions of several optics. $T_\text{f}$, $T_\text{c}^\text{L}$ and $T_\text{c}^\text{F}$ are the transmissions of the optical filter, and the calibration cuvette for the laser and fluorescence signals, respectively.}
					
	\begin{tabular}{| c | cccccccc |}
		\hline	
		Species & $\lambda_\text{L}$ (nm) & $\tau_\text{nat}$ (ns) & $b$ & $E_\text{L}$ ($\upmu$J) &  $T_f$~(\%) & $T_c^L$~(\%) & $T_c^F$~(\%) &  $\eta$~(\%) \\ 	
		\hline	
		O & 225.65 & 34.7 & 1 & 24 & 83.7 & - & - & 9.10 \\
		H & 205.11 & 17.6 & 1 & 35 & 88.9 & - & - & 13.23 \\
		\hline
		Xe & 224.31 & 40.8 & 0.733 & 0.45 & 62.9 & 92 & 94 & 9.65 \\
		Kr & 204.13 & 34.1 & 0.953 & 0.28 & 73.7 & 90 & 94 & 10.35 \\	
		\hline			
	\end{tabular}
					
	\label{tab:TALIF_constants}
					
\end{table}
				
\Cref{tab:TALIF_errors} lists the estimated standard deviations of the individual quantities relevant for the TALIF calibration as well as the overall uncertainty for the absolute atomic density results. The uncertainties of the cross section ratios and natural lifetimes are taken from references~\cite{Niemi2001,Niemi2005}. The uncertainty of $\tau_\text{eff}$ for the calibration gases is calculated as the standard deviation of several independent measurements, as discussed previously.
				
\begin{table}[h]
	\centering
	\small
	\caption[Error estimation for absolute density calibration]{Estimated relative uncertainties (in \%) for several quantities in \cref{eq:TALIF_cal}, as well as the resulting overall relative uncertainties for absolute densities measured under different conditions, which are O densities under an admixture of O$_2$ and H$_2$O, as well as H densities under an admixture of H$_2$O.}
					
	\begin{tabular}{| c| c cccccc c |}
		\hline	
		&& $T_\text{c}$ & $E_\text{L}$ & $\frac{\sigma_\text{cal}^{(2)}}{\sigma_\text{x}^{(2)}}$  & $\tau_\text{nat}$ & $\tau_\text{eff}$ & $S_\text{F}$ & $n$ \\	
		\hline	
		&&&&&&&&\\
		\multirow{4}{*}{\rotatebox[origin=c]{90}{Species}} 
		&O &  & 8 & & 5 &  & - & \\
		&H &  & 8 & & 10 &  & - & \\
		&Xe & 3 & 8 & & 5 & 7 & 10 & \\
		&Kr & 3 & 8 & & 10 & 4 & 10 & \\
		&&&&&&&&\\	
		\hline
		&&&&&&&&\\
		\multirow{5}{*}{\rotatebox[origin=c]{90}{\parbox {1.5cm}{Measurement}}}
		&&&&&&&&\\
		&O-H$_2$O & & & 20 & & 11 & 15 & \textbf{39} \\
		&H-H$_2$O & & & 50 & & 5 & 5 & \textbf{58} \\
		&&&&&&&&\\
		&&&&&&&&\\
		\hline
						
	\end{tabular}
					
	\label{tab:TALIF_errors}
					
\end{table}
				
The resulting uncertainties for the calculation of absolute species densities under several experimental conditions are calculated from the various error bars in \cref{tab:TALIF_errors}. The highest error of 58\% is associated with the measurement of H in H$_2$O containing plasmas due to the high uncertainty of the two-photon excitation cross section ratio.

\section{Global model.}

Measured absolute densities of O and H are compared with a 0D plasma-chemical kinetics model~\cite{Hurlbatt2016} using the GlobalKin code~\cite{Lietz2016}. The considered species and list of reactions are identical to those presented in~\cite{Schroeter2018}. 

The plasma is simulated by assuming a cross section of ($0.1\times0.1$)~cm$^2$ and 3~cm channel length. Simulations are carried out as described in~\cite{Schroeter2018}. The input power is set to 0.3~W, which is in good agreement with~\cite{Golda2016}, and the gas temperature is self-consistently calculated using the GlobalKin code. Simulations are run using a pseudo 1D plug flow and a He gas flow of 500~sccm, resulting in gas velocities around 960~cm/s. For some investigations, simulations are extended into the plasma effluent, where the power is set to zero. In addition, water vapour up to 5000~ppm (0.5\%) is admixed. For some simulations, O$_2$ impurities up to 12~ppm are assumed in the initial gas mixture, corresponding to an air impurity of 60~ppm, in order to investigate the potential effect of gas impurities, i.e. up to 32~ppm air in the used gas bottle of helium, as stated by the supplier, or by reflux from the ambient air into the discharge channel. We do not take into account any nitrogen species in the simulations, however, since we mainly investigate the formation of oxygen and hydrogen containing species in this work, we assume that nitrogen species are not directly involved in their formation mechanisms. 

\section{Determination of quenching with H$_2$O.}

\begin{figure}[ht]  
	\centering
	
	\includegraphics[width=8cm]{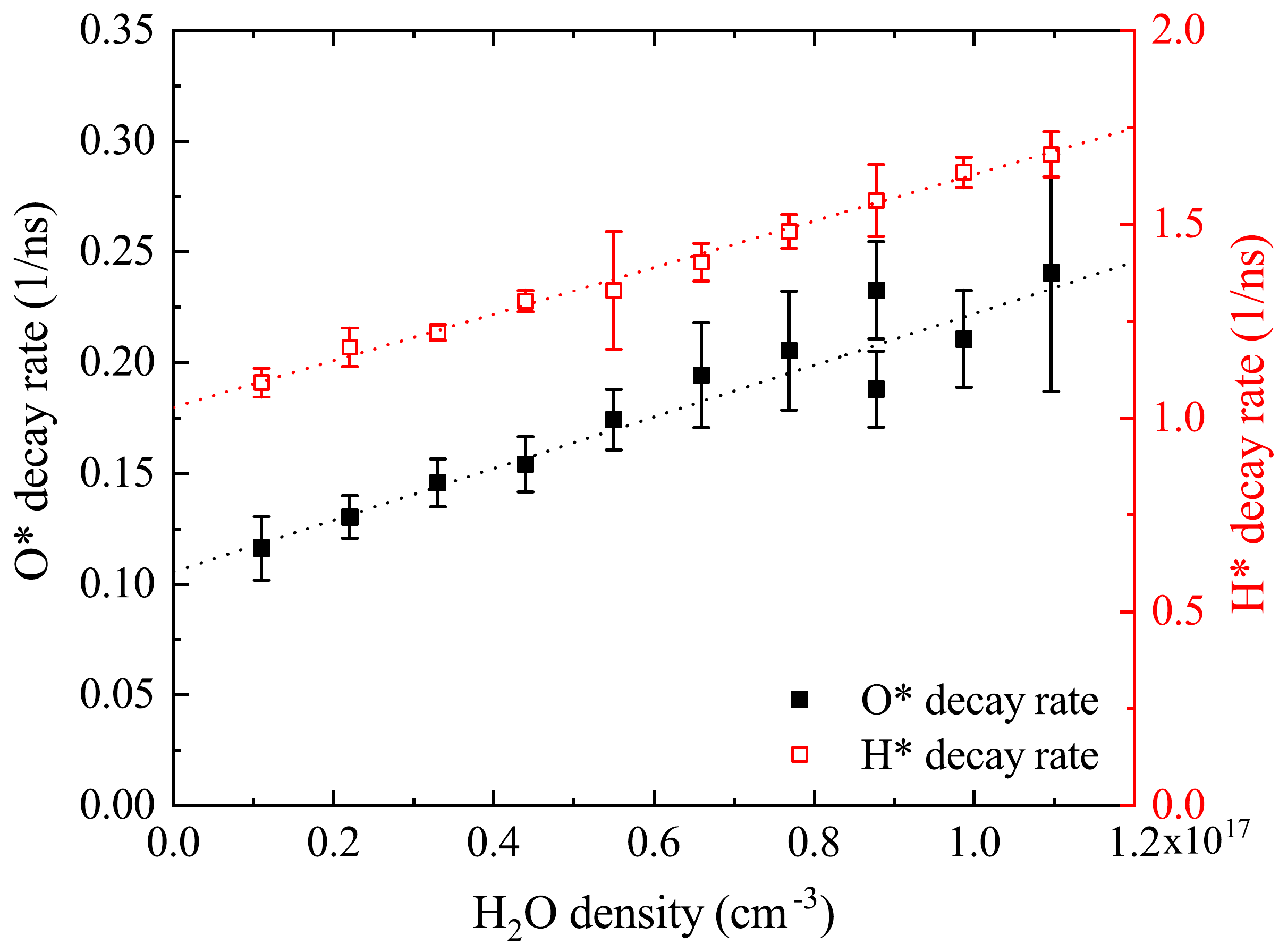}
	
	\caption{Measured decay frequencies of O$^*$ and H$^*$ as a function of H$_2$O content in the feed gas. Measurements were taken at 500~sccm total gas flow, and 510~V$_\text{pp}$. Error bars show uncertainties from the exponential decay fitting to the raw data.}
	\label{fig:Quenching_h2o}
\end{figure}

The sub-nanosecond temporal resolution of the experimental diagnostic setup enables the measurement of effective decay rates (natural decay rate and influence of collisional quenching) at atmospheric pressure. This allows for the determination of quenching coefficients for the laser-excited states with several quenching molecules, in this case H$_2$O. For this, effective decay rates are measured as a function of H$_2$O density in the feed gas, which is shown in \cref{fig:Quenching_h2o} for O$^*$ and H$^*$. If the decay rates are linearly dependent on the H$_2$O content, quenching coefficients can be calculated from the slopes of the linear fits. \Cref{tab:Lit_quench} shows a comparison of quenching coefficients obtained in this work with literature values. They will be discussed in more detail in the next sections.

\begin{table}[h]
	\centering
	\caption{Measured quenching coefficients for O$^*$ and H$^*$ with H$_2$O and comparison with literature values.}
	
	\begin{tabular}{| c c | ccc |}
		\hline	
		Species & Quenching species & $k_\text{q}$~(cm$^3$s$^{-1}$)  & $k_\text{q}^\text{lit}$~(cm$^3$s$^{-1}$) & Ref.  \\ 	
		\hline	
		O(3p $^3$P) & H$_2$O & $1.1\times10^{-9}$ ($\pm10$\%) & $(9.4\pm1.5)\times10^{-10}$ & \cite{Quickenden1979} \\
		&&& $(4.9\pm0.3)\times10^{-9}$ & \cite{Meier1986} \\
		\hline
		H(n=3) & H$_2$O & $6.0\times10^{-9}$ ($\pm3$\%)& $(9.1\pm1.6)\times10^{-9}$ & \cite{Quickenden1979}   \\
		&&& $(1.1\pm0.1)\times10^{-8}$ & \cite{Meier1986} \\	
		\hline			
	\end{tabular}
	
	\label{tab:Lit_quench}
	
\end{table}

When calculating the quenching coefficients from the slopes of the fits in \cref{fig:Quenching_h2o}, it is assumed that the majority species He and H$_2$O in the feed gas are the main quenching partners, while disregarding the fraction of H$_2$O that is dissociated $D_{\text{H}{_2}\text{O}}$~(\%) 
\begin{equation}
D_{\text{H}_2\text{O}} = \frac{n_{\text{H}_2\text{O}}(0\text{ cm})-n_{\text{H}_2\text{O}}(3\text{ cm})}{n_{\text{H}_2\text{O}}(0\text{ cm})}\times 100\:.
\end{equation}

At $P=0.3$~W, at very low H$_2$O content in the feed gas (1~ppm), up to 33\% of the admixed H$_2$O molecules are dissociated in the plasma by the time they reach the end of the plasma channel. The dissociation fraction rapidly decreases with increasing H$_2$O admixture. For admixtures greater than 500~ppm, the dissociation degree settles to a value of about 3.8\%, and provides an additional uncertainty to the values in \cref{tab:Lit_quench}.

A fluctuation of the bubbler temperature within the error bar of 1$^\circ$C leads to an uncertainty of 6\%, in addition to the uncertainties in \cref{tab:Lit_quench}. The bubbler is not temperature controlled for these measurements.

\subsection{Quenching of O(3p $^3$P) with H$_2$O}\label{sec:H2O_quench}

The quenching coefficient $k_{\text{H}_2\text{O}}^{\text{O}^*}=1.1\times10^{-9}$~cm$^3$s$^{-1}$ is obtained from the slope of measured decay rates shown in \cref{fig:Quenching_h2o} (black squares and dashed black line) with an uncertainty of 10\% from the linear fit.

A literature value for the quenching coefficient of O$^*$ with H$_2$O has been obtained previously by Quickenden et al.~\cite{Quickenden1979} as  $k_{\text{H}_2\text{O}}^{\text{O}^*}=(9.4\pm1.5)\times10^{-10}$~cm$^3$s$^{-1}$ using radiolysis of pure water vapour with an electron beam, and detection of fluorescence light using a photo-multiplier. The water vapour was created by heating up a supply of water connected to their experimentation cell. Meier et al.~\cite{Meier1986} have measured the same quenching coefficient using TALIF in a flow-tube reactor as $k_{\text{H}_2\text{O}}^{\text{O}^*}=(4.9\pm0.3)\times10^{-9}$~cm$^3$s$^{-1}$ under low pressure conditions in the mbar range. The literature values for quenching coefficients for O$^*$ with H$_2$O do not agree well with each other. The O$^*$ quenching coefficient with H$_2$O obtained in this work lies between the two available literature values, but closer to the quenching coefficient measured by Quickenden et al.~\cite{Quickenden1979}, although this is the older of the two cited sources, and a non-linearity between quenching rates and H$_2$O content was observed in their experiment.

By extrapolating the linear fit to zero H$_2$O admixture in \cref{fig:Quenching_h2o}, a lifetime of $\tau_{\text{O}^*\text{,eff}}=9.6$~ns is obtained for pure He. Taking the natural lifetime of O$^*$ $\tau_{\text{O}^*\text{,nat}}=34.7$~ns~\cite{Niemi2005}, a quenching coefficient for O$^*$ with He can be determined as $k_{\text{He}}^{\text{O}^*}=0.032\times10^{-10}$~cm$^3$s$^{-1}$ using
\begin{equation}
k_{\text{He}}^{\text{O}^*}=\frac{\frac{1}{\tau_{\text{O}^*\text{,eff}}}-\frac{1}{\tau_{\text{O}^*,nat}}}{n_\text{He}}\:.\label{eq:quench}
\end{equation} 
This value lies within the broad span of the literature values ranging from 0.016 to $0.15\times10^{-10}$~cm$^3$s$^{-1}$~\cite{Schmidt2016,Niemi2005,Niemi2001,Bittner1988}.

\subsection{Quenching of H(n=3) with H$_2$O.}

The quenching coefficient $k_{\text{H}_2\text{O}}^{\text{H}^*}=6.0\times10^{-9}$~cm$^3$s$^{-1}$ is obtained from \cref{fig:Quenching_h2o} (red triangles) with an uncertainty of 3\% from the linear fit. The measured values are significantly smaller than the literature values $k_{\text{H}_2\text{O}}^{\text{H}^*}=(9.1\pm1.6)\times10^{-9}$~cm$^3$s$^{-1}$ and $k_{\text{H}_2\text{O}}^{\text{H}^*}=(1.1\pm0.1)\times10^{-8}$~cm$^3$s$^{-1}$ from~\cite{Quickenden1979,Meier1986}, respectively, although obtained by the same techniques as described in \cref{sec:H2O_quench}.

From the intercept of linear fit in \cref{fig:Quenching_h2o}, an effective lifetime $\tau_{\text{H}^*\text{,eff}}=1.0$~ns is obtained in pure He. Using the literature value $\tau_{\text{H}^*\text{,nat}}=17.6$~ns~\cite{Schmidt2016,Niemi2001,Preppernau1995} for the natural lifetime of H$^*$ and \cref{eq:quench}, we derive the quenching coefficient $k_{\text{He}}^{\text{H}^*}=0.42\times10^{-10}$~cm$^3$s$^{-1}$. This value is close to the most recently determined literature value $k_{\text{He}}^{\text{H}^*}=0.317\times10^{-10}$~cm$^3$s$^{-1}$~\cite{Schmidt2016}, while older literature values range from 0.099 to $0.53\times10^{-10}$~cm$^3$s$^{-1}$~\cite{Bittner1988,Niemi2001,Preppernau1995}. 

\section{Atomic species as a function of humidity.}

\subsection{Atomic oxygen.}

\begin{figure}[ht]  
	\centering
	
	\begin{subfigure}[h]{0.455\linewidth}
		\includegraphics[width=\textwidth]{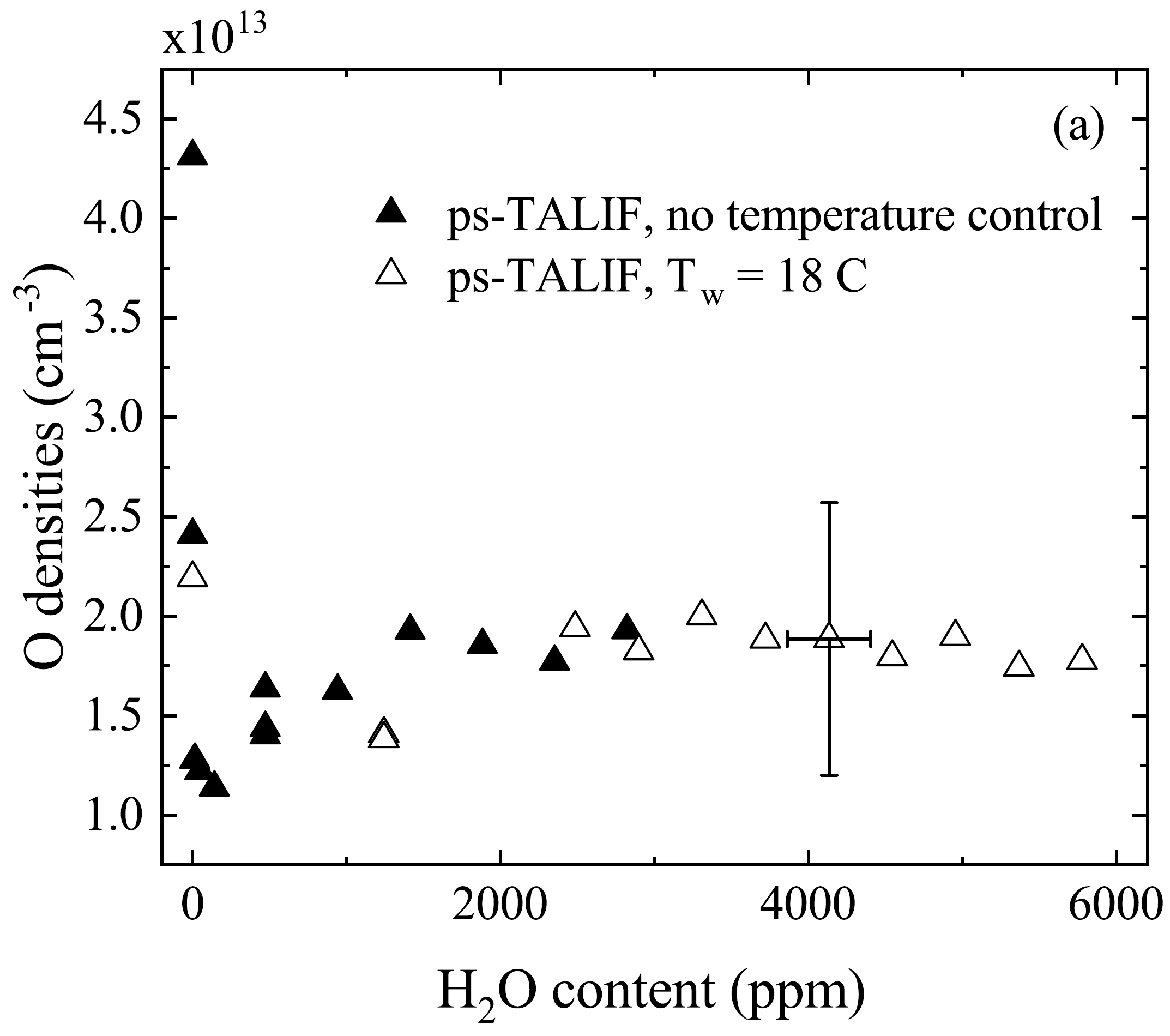}	
	\end{subfigure}
	\hfill
	\begin{subfigure}[h]{0.51\linewidth}
		\includegraphics[width=\textwidth]{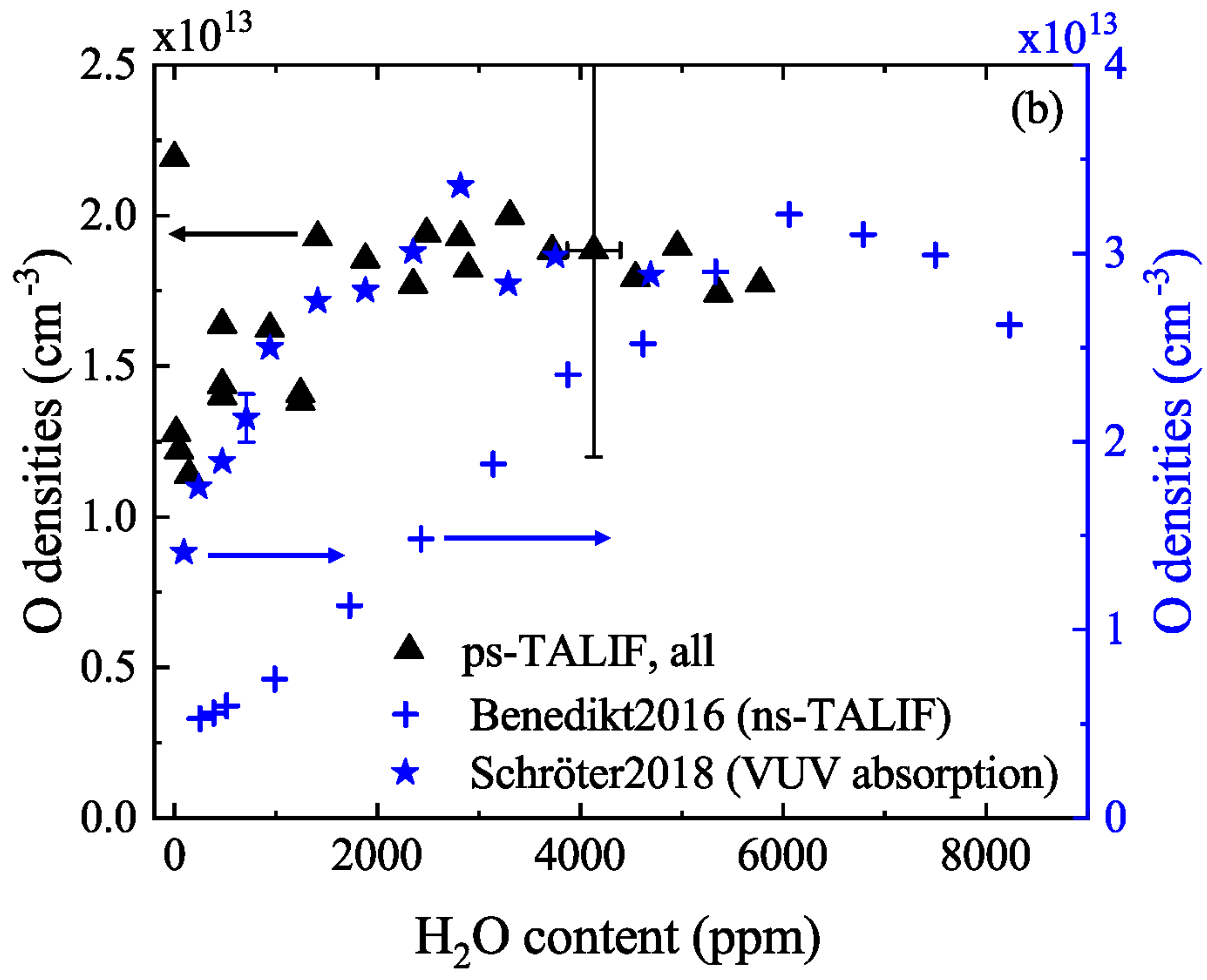}
	\end{subfigure}
	\caption{Absolute ground state atomic oxygen densities as a function of humidity content of the He feed gas (black triangles). Measurements were taken at 500~sccm total gas flow, and 510~V$_\text{pp}$. (a) ps-TALIF measurements. (b) Zoom into ps-TALIF results, ignoring high density values at low H$_2$O admixtures. Data shown in blue is a comparison with literature values. Stars: Schr\"{o}ter et al.~\cite{Schroeter2018}; crosses: Benedikt et al.~\cite{Benedikt2016}. }
	\label{fig:O_watervar_lit}
\end{figure}

\Cref{fig:O_watervar_lit}~(a) shows the O densities produced under a variation of humidity in the feed gas. Above 100~ppm, the O density increases slightly with increasing humidity admixtures, until it reaches a maximum level of about $2\times10^{13}$~cm$^{-3}$ at around 2000-2500~ppm. At very low H$_2$O content, however, a sharp increase of O densities with decreasing H$_2$O content is observed, which is shown in \cref{fig:O_watervar_lit}~(a). A peak value of $4.3\times10^{13}$~cm$^{-3}$ is obtained before H$_2$O is actively admixed to the He background gas. Interestingly, this high value can not be reproduced when this data point is retaken after admixing H$_2$O to the feed gas. This is a strong indicator that residual H$_2$O from previous measurements attached to the feed gas line is still present, which changes the plasma chemistry, compared to a measurement that was started with a dry feed gas line. Therefore, the points measured at low \textit{intentional} H$_2$O admixtures are believed to be very dependent on feed gas impurities, as discussed further below.

The observed trend agrees very well with previously investigated O densities measured by Vacuum Ultra-Violet Fourier-Transform Absorption Spectroscopy (VUV-FTAS)~\cite{Schroeter2018}, which are shown in \cref{fig:O_watervar_lit}~(b) as blue stars. However, maximum O densities measured with ps-TALIF in the $\upmu$APPJ are about 70\% of the O densities measured with VUV-FTAS in previous investigations. Possible reasons for this difference could lie in the different surface-to-volume ratios of the two sources, resulting in different recombination probabilities for species such as H at the reactor walls~\cite{Schroter2018}, leading to a changed chemistry in the plasma bulk, and the fact that O is measured at different positions in the jet using the two setups (in the centre of the discharge at 1.2~cm in \cite{Schroeter2018} and outside the channel in this work). Differences in measured absolute densities can also arise from the two very different diagnostic techniques used here and in the previous investigations (ps-TALIF vs. VUV-FTAS) and the uncertainties associated with them.

A comparison with the data of Benedikt et al.~\cite{Benedikt2016}, also shown in \cref{fig:O_watervar_lit}~(b) as blue crosses, show good quantitative agreement with our results within a factor 1.5--2.5, depending on H$_2$O content. Their measurements have been carried out using ns-TALIF in a controlled He atmosphere. Under their plasma operating conditions (1.4~slm total He flow, 565~V$_\text{pp}$), they observed an increase in O density up to $3\times10^{13}$~cm$^{-3}$ at 6000~ppm, followed by a short decrease until the plasma extinguished at 8000~ppm. Therefore, their measured O peak densities are about a factor 2.5 larger than the values obtained in this work. One possible reason for the difference lies in the fact that with their ns-TALIF system, Benedikt et al. were not able to determine their O$^*$ lifetimes experimentally. Instead, they calculated the effective decay rate using the quenching coefficient $k_{\text{H}_2\text{O}}^{\text{O}^*}=4.9\times10^{-9}$~cm$^{3}$s$^{-1}$ from reference~\cite{Meier1986}, which is much larger than the coefficient $k_{\text{H}_2\text{O}}^{\text{O}^*}=1.3\times10^{-9}$~cm$^3$s$^{-1}$ obtained in this work. Since the absolute O densities depend linearly on $A_\text{eff}$, as shown in \cref{eq:TALIF_cal_short}, and therefore on the quenching coefficient, the use of a larger quenching coefficient would lead to higher densities, and potentially also different trends in O densities, as observed in \cref{fig:O_watervar_lit}. In addition, our experimental investigations were carried out under slightly different plasma conditions (510~V$_\text{pp}$, 0.5~slm He flow) compared to the investigations by Benedikt et al. (565~V$_\text{pp}$, 1.4~slm He flow), which could also affect both observed trends and absolute densities. Particularly the higher applied voltage in the work of Benedikt et al. could lead to a shift of maximum O densities towards higher water contents.

Additionally, using ps-TALIF, a strong increase of O towards very low admixtures is observed in this work, which can be seen best in \cref{fig:O_watervar_lit}~(a). These high O densities at low admixtures are most likely due to impurities in the feed gas, either because of O$_2$ entering the feed gas through small leaks, or diffusion of O$_2$ from the ambient air. This assumption is supported by the fact that we did not observe this in earlier work using VUV-FTAS~\cite{Schroeter2018}, where measurements were carried out in a closed source with no contact to ambient air. In addition, Benedikt et al.~\cite{Benedikt2016} do also not observe these high densities at low admixtures, quite likely because their measurements were carried out in a pure He atmosphere. In addition, we assume that this effect would be less observable at higher flow rates as used by Benedikt et al., because ambient gas is less likely to enter the plasma jet.

\begin{figure}[ht]  
	\begin{subfigure}[h]{0.48\linewidth}
		\includegraphics[width=\textwidth]{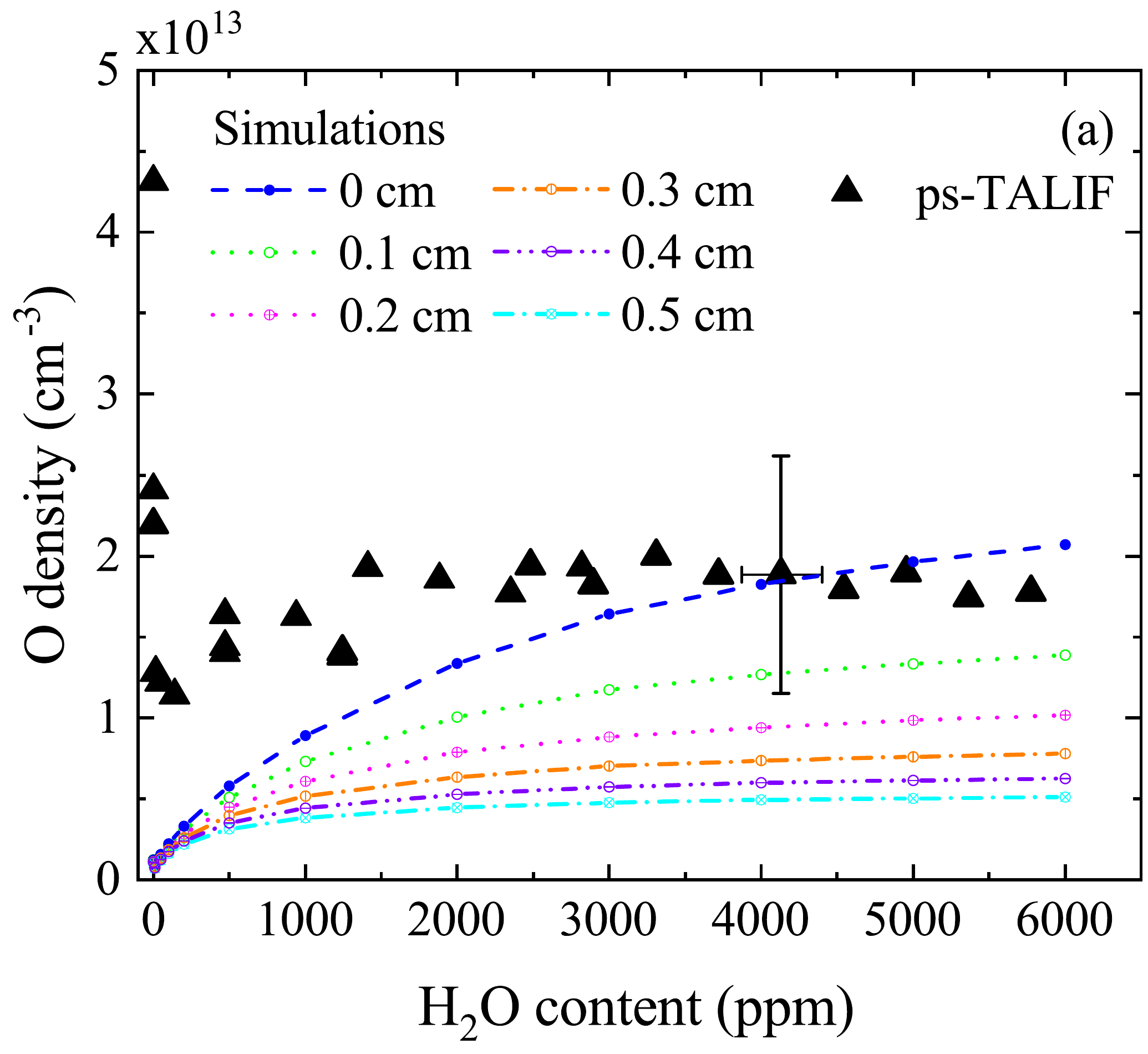}
	\end{subfigure}
	\hfill
	\begin{subfigure}[h]{0.48\linewidth}
		\includegraphics[width=\textwidth]{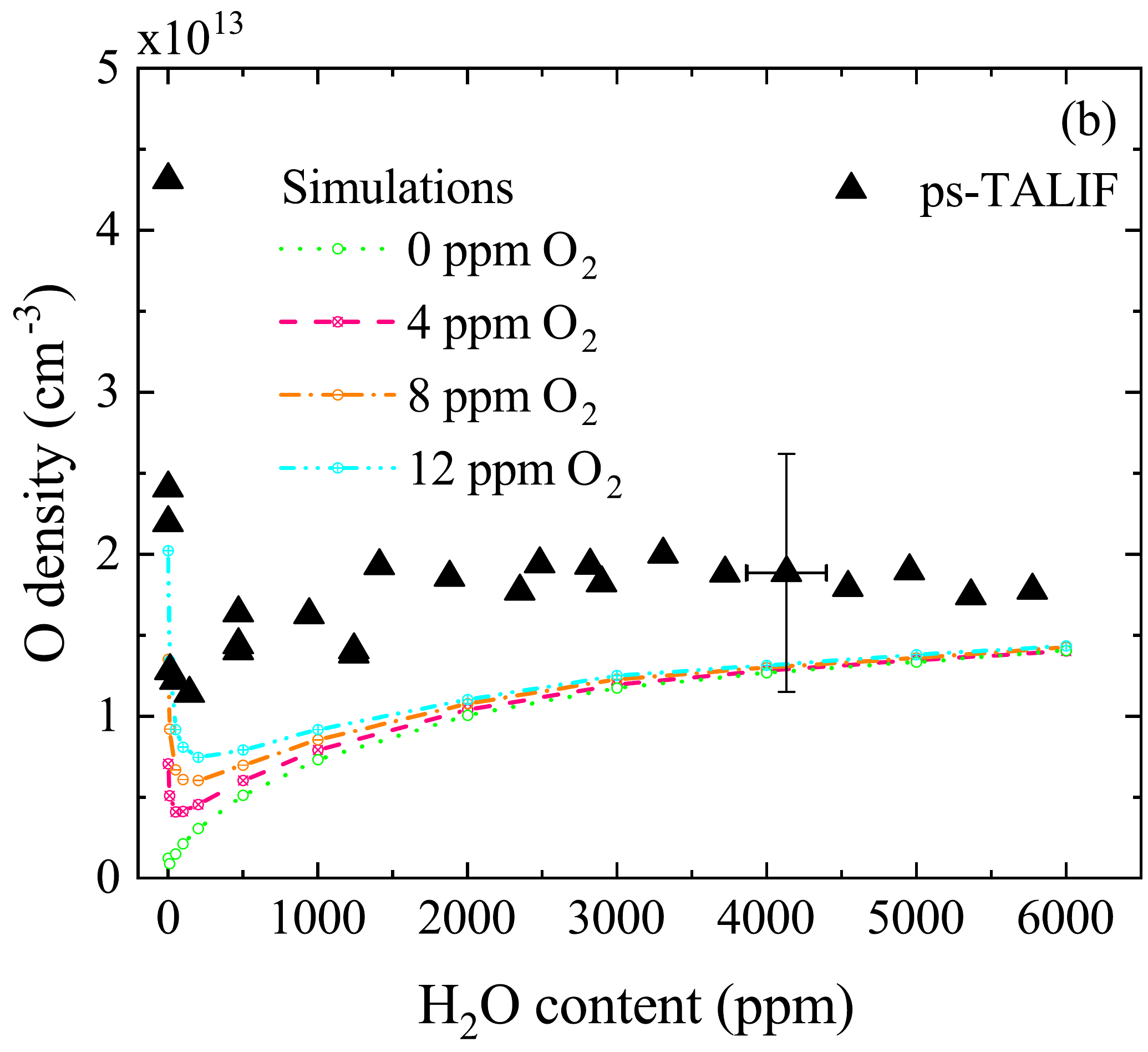}
	\end{subfigure}\centering
	
	\caption{Measured absolute O densities under a variation of the humidity content (triangles), as in \cref{fig:O_watervar_lit}. (a) Simulation results for absolute O densities as a function of H$_2$O content and distance from the plasma nozzle, for a plasma power of 0.3~W. (b) O densities as a function of H$_2$O for different air impurity contents, for a plasma power of 0.3~W and distance 0.1~cm from the nozzle.}
	\label{fig:APPJ_O_watervar}
\end{figure}

In order to investigate the influence of impurities and distance to the plasma nozzle further, GlobalKin simulations are carried out for the $\upmu$APPJ. \Cref{fig:APPJ_O_watervar} shows simulated absolute atomic oxygen densities together with the measurement results for a variation of the humidity content in the feed gas.

The dashed lines in \cref{fig:APPJ_O_watervar}~(a) indicate the absolute simulated O densities as a function of H$_2$O content in the feed gas, for different distances from the nozzle, and without any air impurities. The O density decreases with increasing distance to the nozzle, due to consumption of O in chemical reactions, mainly with OH 
\begin{equation}
O + OH \rightarrow O_2 + H \:.\label{eq:O_OH}
\end{equation}
Consumption by reactions with HO$_2$ gain more importance further away from the nozzle
\begin{equation}
O + HO_2 \rightarrow O_2 + OH \:.\label{eq:O_HO2}
\end{equation}
The trends for O densities under a humidity variation depend on the distance from the jet nozzle. While directly at the nozzle at 0~cm (blue line in \cref{fig:APPJ_O_watervar}~(a)), O densities are monotonically increasing with increasing H$_2$O admixture, O densities a few millimetres away from the nozzle approach a steady-state value at high H$_2$O admixtures. ps-TALIF measurements were carried out at approximately 1~mm distance to the plasma nozzle (green line in \cref{fig:APPJ_O_watervar}~(a)). At this distance, trends in the simulation and experiments are slightly different at high H$_2$O content. In the simulation, O densities clearly increase with increasing H$_2$O admixture. In the experiment, a plateau is reached at about 1500~sccm. At higher admixtures, O densities stay constant within the error bars of the measurement. Simulated and measured absolute O densities agree within about a factor of 2, particularly considering that the formation of O from H$_2$O is complex, in the sense that it is not primarily produced from direct dissociation of H$_2$O molecules.

In order to investigate the role of impurities in the feed gas flow, different O$_2$ impurity concentrations of 4, 8, and 12~ppm are assumed in the initial gas mixture. As a comparison, the amount of air impurity in a He 4.6 grade bottle is 32~ppm according to the gas supplier, corresponding to an O$_2$ impurity of 6.4~ppm. The absolute O densities simulated under these conditions are shown in \cref{fig:APPJ_O_watervar}~(b) for a distance of 1~mm from the nozzle. In the simulation, at very low H$_2$O admixtures, $<$~100~ppm, O densities sharply increase towards decreasing H$_2$O admixtures, as observed in the experiments. It is therefore likely that the trend in the experiment is due to air impurities being present in the feed gas. At high humidity admixtures, the addition of air impurities only makes a small difference to absolute O densities. It is therefore concluded that plasmas can be operated in a more controlled way by purposefully admixing molecules into the feed gas, because the produced RS are not as strongly influenced by ambient conditions, which are susceptible to change unless the plasma is operated in a shielding gas atmosphere.

\begin{figure}[h!]  

	\includegraphics[width=\textwidth]{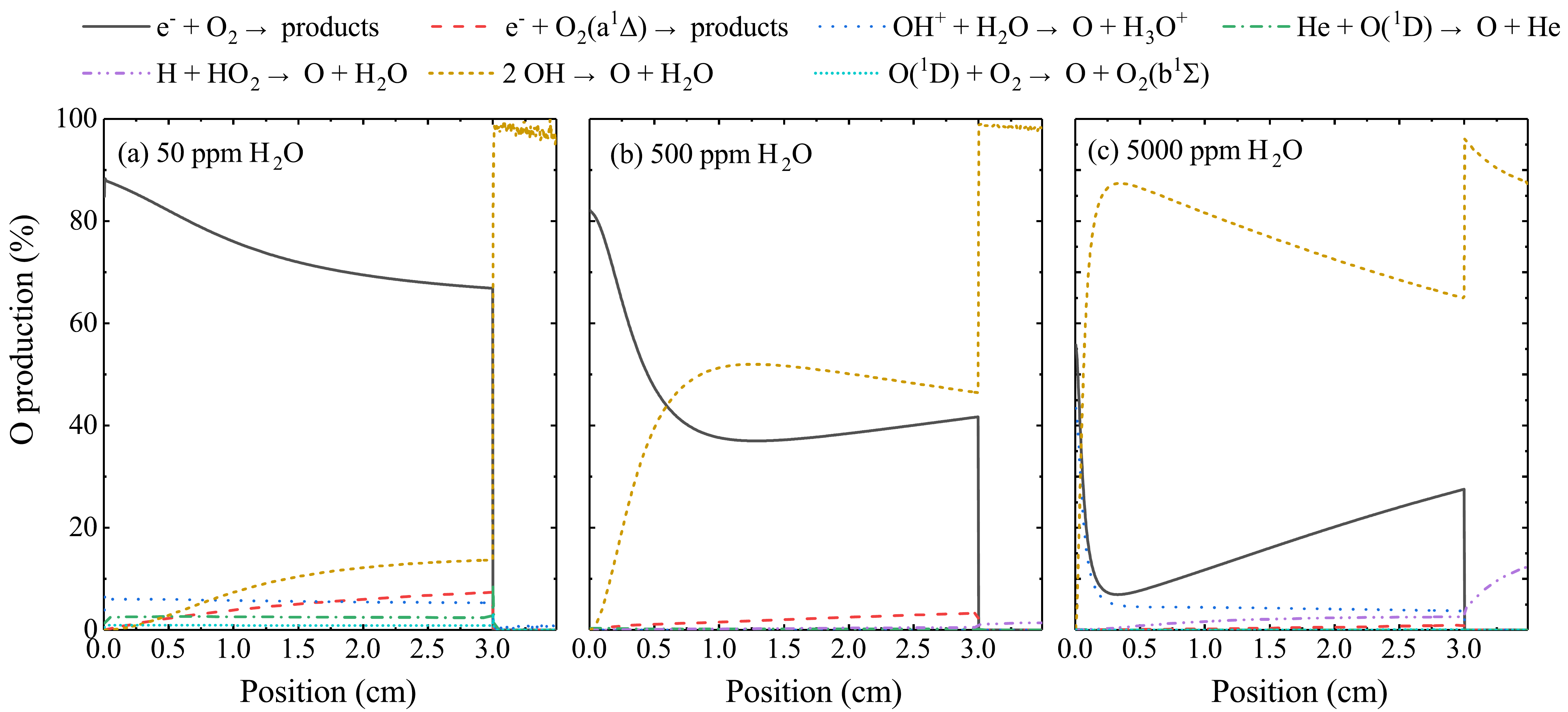}	

	\caption[Formation pathways for O at different humidity contents]{Formation pathways for O at different humidity contents for a fixed O$_2$ admixture of 8~ppm, 0.3~W and 500~sccm total He flow.}
	\label{fig:APPJ_O_production}
\end{figure}

\begin{figure}[h!]  

	\includegraphics[width=\textwidth]{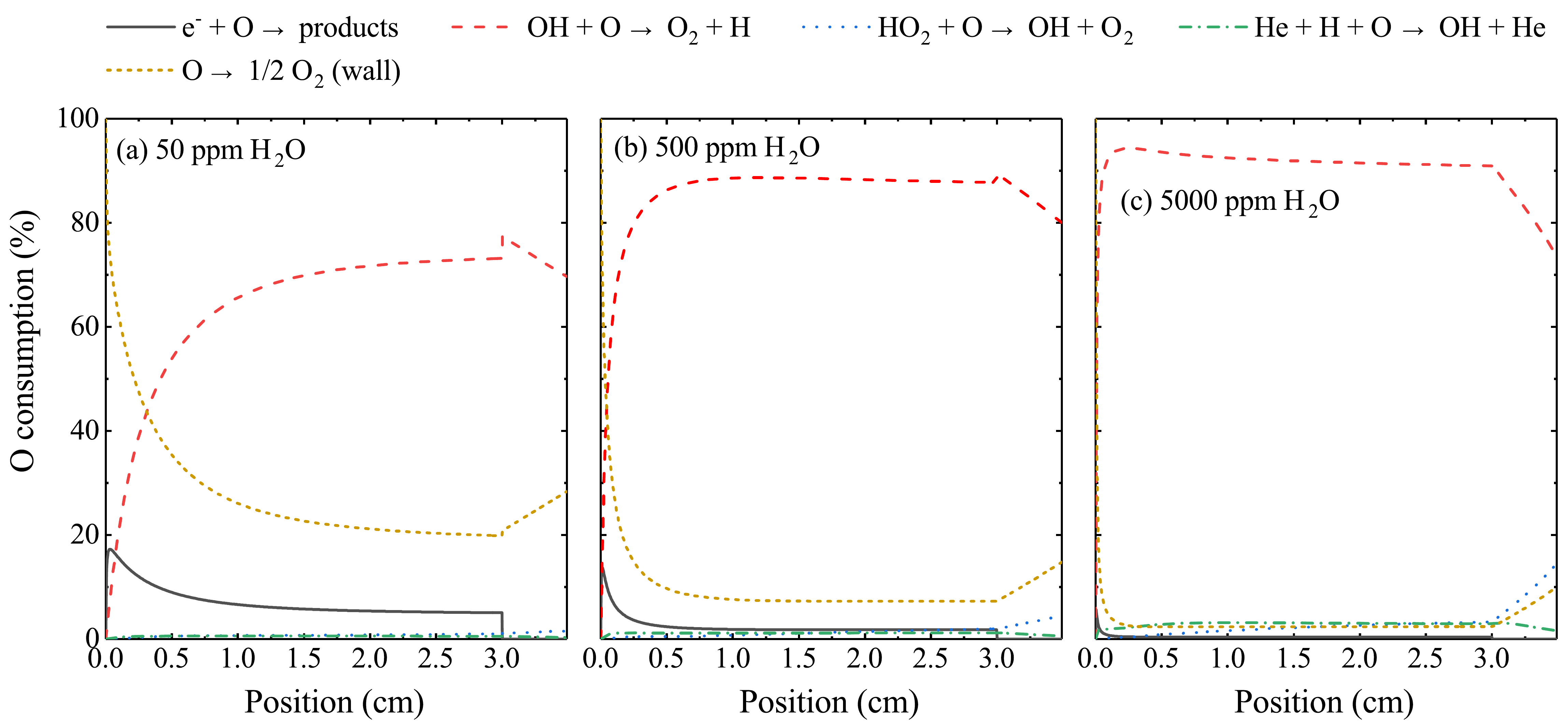}
	
	\caption[Consumption pathways for O at different humidity contents]{Consumption pathways for O at different humidity contents and a fixed O$_2$ admixture of 8~ppm, 0.3~W and 500~sccm total He flow.}
	\label{fig:APPJ_O_consumption}
\end{figure}

The very good agreement between simulations and experiments means that the production and consumption reactions for different H$_2$O admixtures can be investigated. In the following discussion, a pathway analysis is carried out for an O$_2$ impurity content of 8~ppm because this value is closest to the intrinsic oxygen impurity level of the used He feed gas. An overview of the different pathways can be found in \cref{fig:APPJ_O_production,fig:APPJ_O_consumption}.

The production pathways for O at different H$_2$O contents are shown in \cref{fig:APPJ_O_production}. At a H$_2$O admixture of only 50~ppm, the plasma chemistry is mainly dominated by the oxygen reactions originating from oxygen impurities. O is mainly produced by electron impact dissociation of molecular oxygen
\begin{align}
e + O_2 &\rightarrow O + O(^1D) + e \label{eq:dissO21}\:,\\
e + O_2 &\rightarrow 2O + e \:,\label{eq:dissO22}
\end{align}
and the equivalent processes from the O$_2$(a$^1\Delta$) and O$_2$(b$^1\Sigma$) states. Further O is produced via quenching of excited O($^1$D)
\begin{equation}
O(^1D) + O_2 \rightarrow O + O_2\:. \label{eq:dissO23} 
\end{equation}
\Cref{eq:dissO21,eq:dissO22,eq:dissO23} have been previously identified as being the main production channels for O in He-O$_2$ APPs in a similar plasma source~\cite{Waskoenig2010b}.

As the H$_2$O content increases, these processes become less relevant, and are increasingly replaced by reactions including hydrogen containing species. Particularly 
\begin{equation}
2OH \rightarrow H_2O + O \:,
\end{equation}
which was found to be important for the formation of O in~\cite{Schroeter2018} and is found to play a key role in the formation of O under these conditions as well. This reaction becomes the dominant production mechanism at high H$_2$O admixtures. Dissociation of O$_2$ remains one of the dominant production mechanisms also at high H$_2$O admixtures, since O$_2$ is actively formed in H$_2$O containing plasma, as discussed in~\cite{Schroeter2018}. On the other hand, densities of O($^1$D) decrease rapidly with increasing admixture of water, therefore, quenching of O($^1$D) becomes less important for the production of O at higher H$_2$O contents.

The O consumption pathways are shown in \cref{fig:APPJ_O_consumption}. For 50 ppm H$_2$O, the first few mm in the plasma are dominated by losses to the wall. After that, and for all positions for higher H$_2$O content, the main consumption pathway is \cref{eq:O_OH},
%
\begin{equation}
OH + O \rightarrow O_2 + H . \notag
\end{equation}
%
%


\subsection{Atomic hydrogen.}

\begin{figure}[h!]  
	\centering
	
	\begin{subfigure}[t]{0.48\linewidth}
		\includegraphics[width=\textwidth]{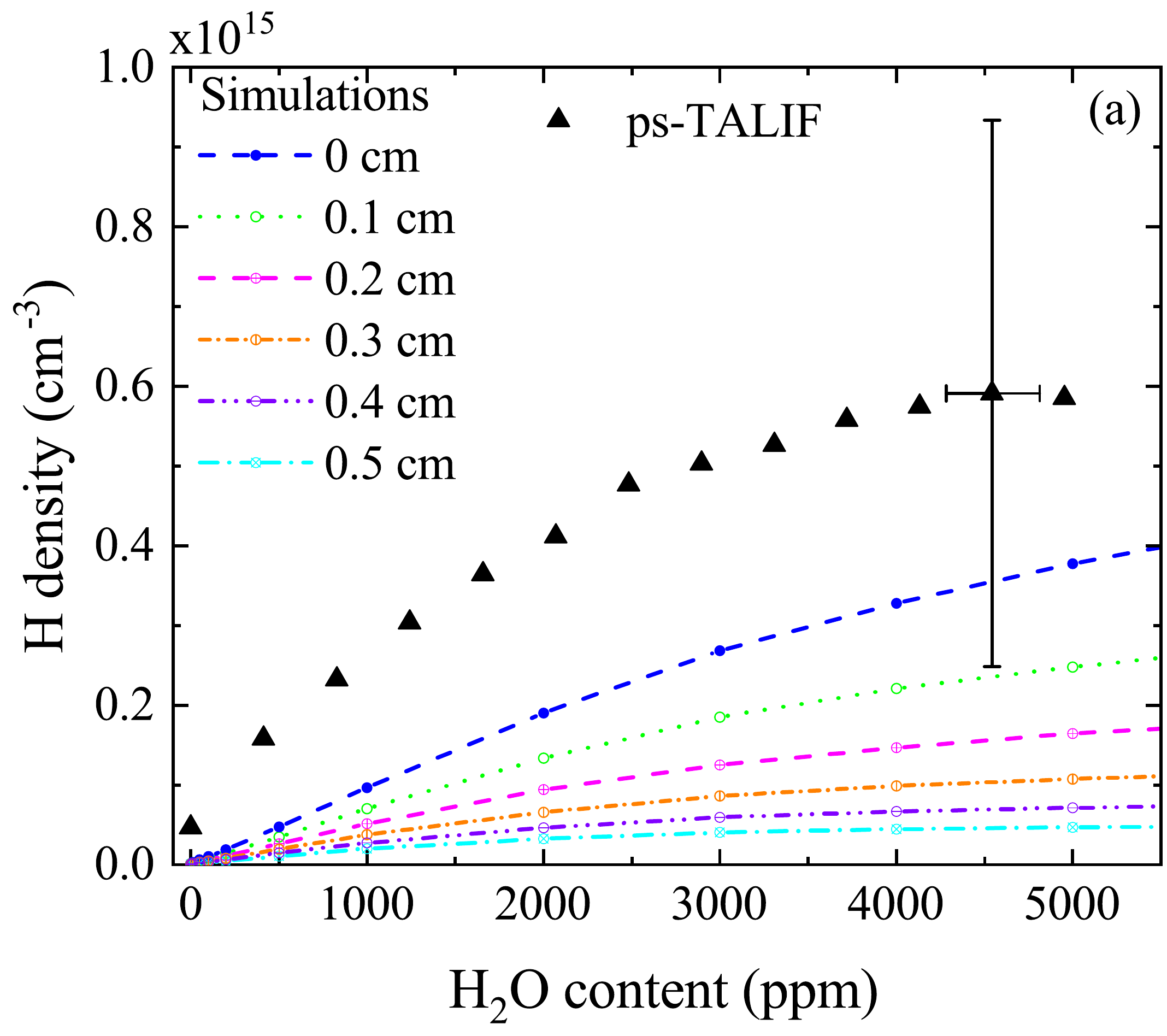}	
	\end{subfigure}
	\hfill
	\begin{subfigure}[t]{0.48\linewidth}
		\includegraphics[width=\textwidth]{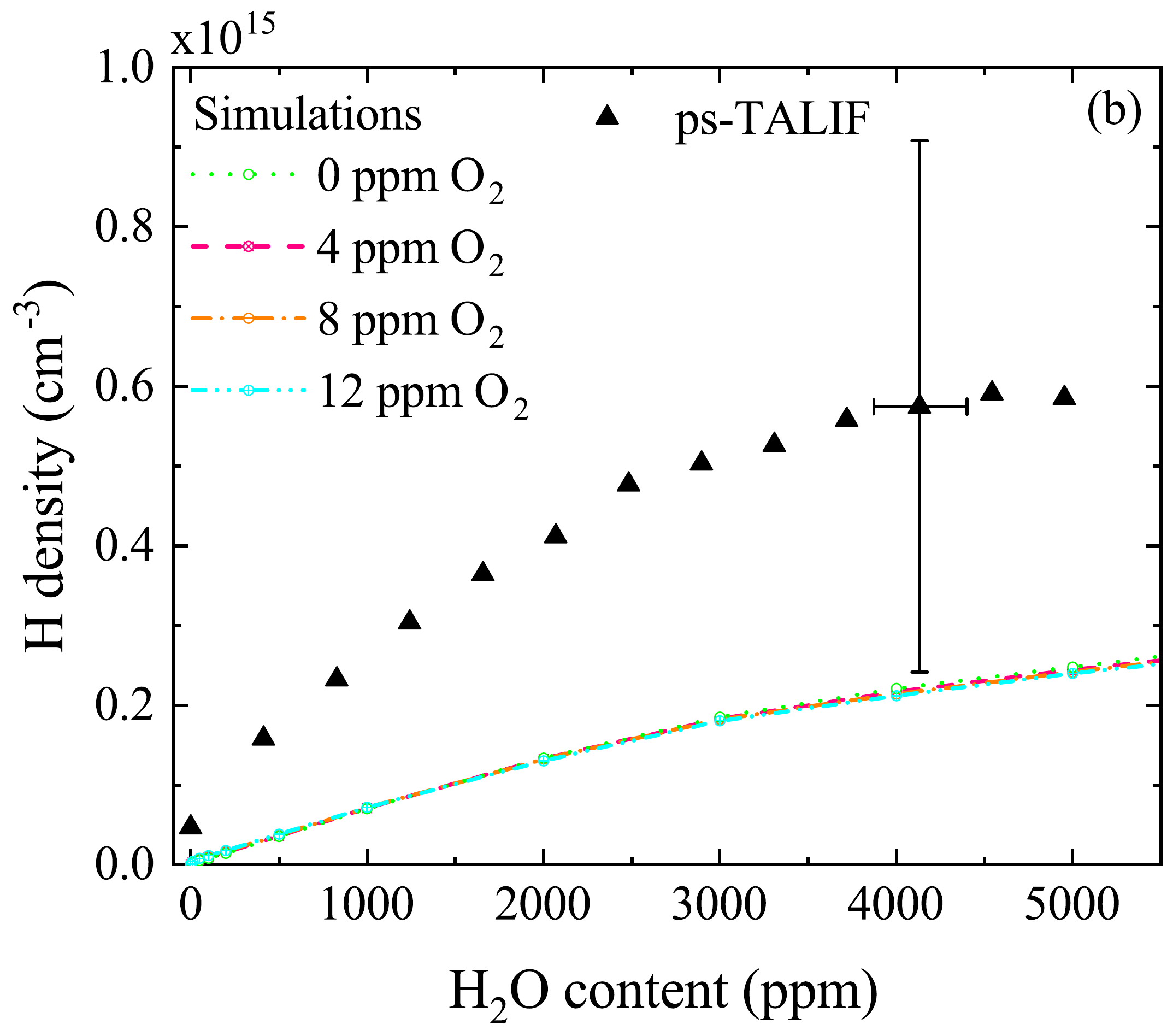}
	\end{subfigure}
	
	\caption[Experimental and numerical H densities as a function of gas humidity content]{Absolute ground state atomic hydrogen densities as a function of gas humidity content (black triangles). Also shown are simulation results for different distances to the jet nozzle (a), and different air impurity contents (b). Measurements were taken at 500~sccm total gas flow, and 510~V$_\text{pp}$. The input power for the simulations is 0.3~W.}
	\label{fig:H_watervar}
\end{figure}

\Cref{fig:H_watervar} shows absolute H densities as a function of H$_2$O admixture in the feed gas (black triangles). H densities increase monotonically with increasing humidity content over the whole measurement range. 

Similar to \cref{fig:APPJ_O_watervar} for atomic oxygen, the dependence of H densities on the distance to the plasma jet nozzle and air impurity content is investigated. From \cref{fig:H_watervar}~(a) can be seen that the simulations are predict lower H densities for all distances. However, simulations of H densities for distances of 0 and 0.1~cm from the plasma jet are closest to the measured values, and within the estimated experimental error. As discussed previously, these distances match best the distance measured in the experiment. Therefore, experiment and simulations are in reasonable agreement. \Cref{fig:H_watervar}~(b) shows H densities for different oxygen impurity contents in the feed gas. From the results it is clear that H densities are almost independent on the amount of impurities in the plasma, in contrast to the trends found for O, where impurities had a large influence on O densities at low water admixtures.

\begin{figure}[h!]  
	
	\includegraphics[width=\textwidth]{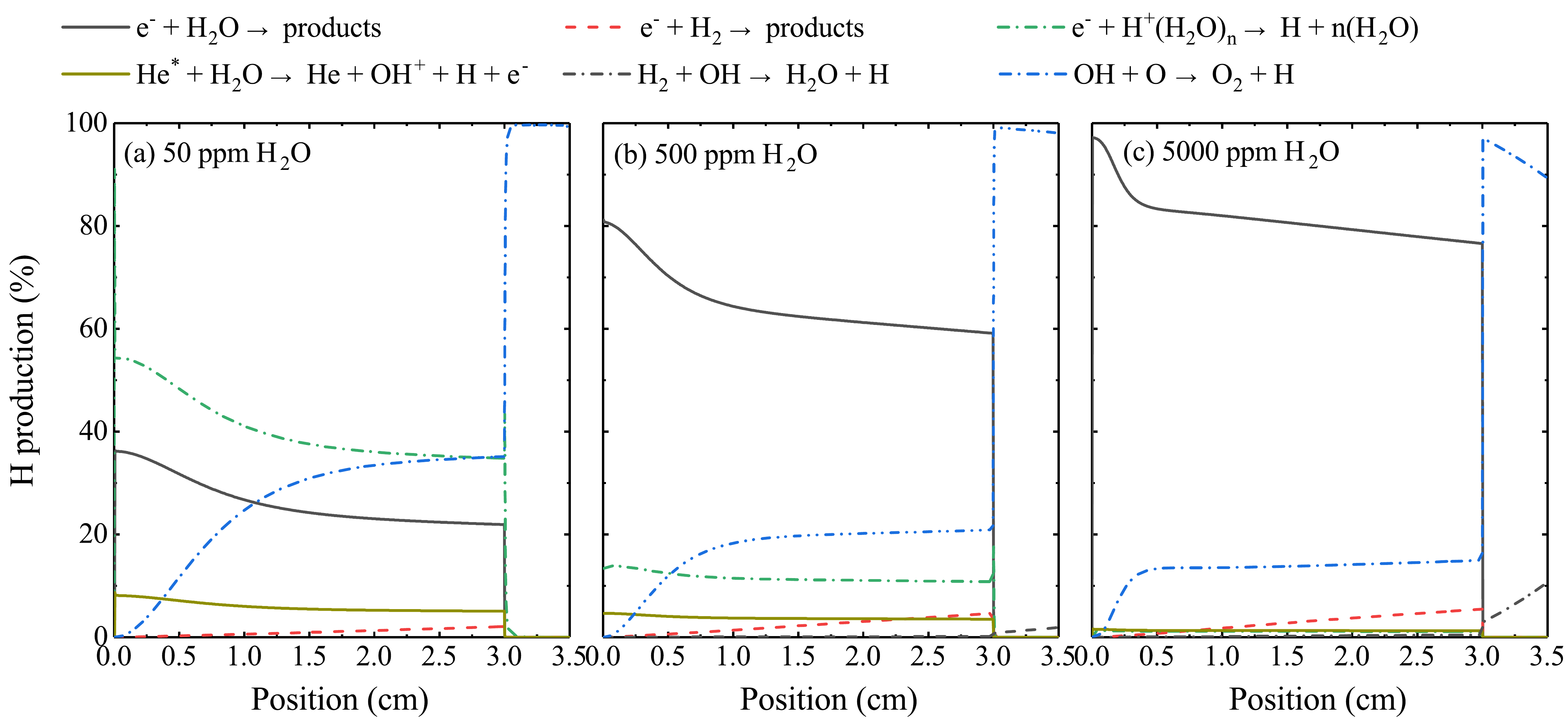}	
	
	\caption[Formation pathways for H at different humidity contents]{Formation pathways for H at different humidity contents and a fixed O$_2$ admixture of 8~ppm, 0.3~W and 500~sccm total He flow.}
	\label{fig:APPJ_H_production}
\end{figure}

\begin{figure}[h!]  
	
	\includegraphics[width=\textwidth]{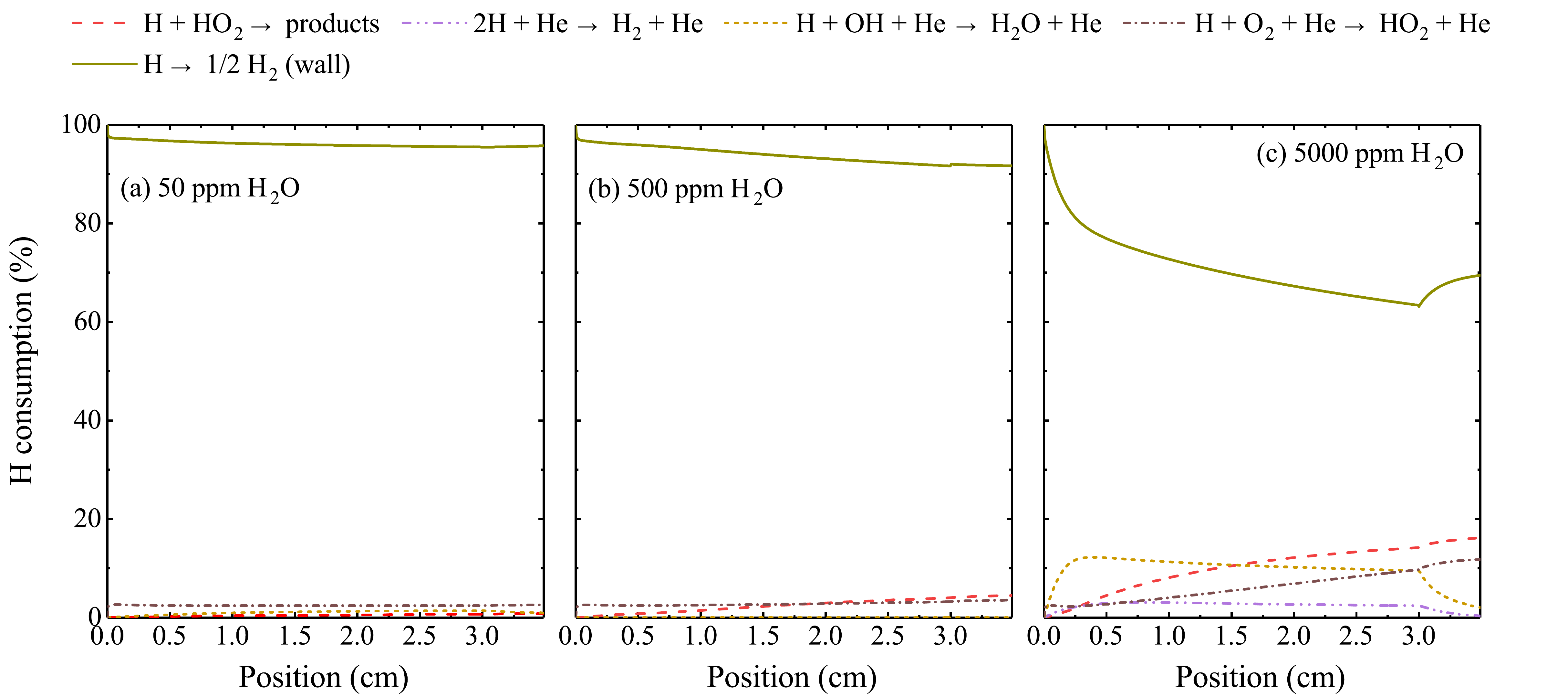}
	
	\caption[Consumption pathways for H at different humidity contents]{Consumption pathways for H at different humidity contents and a fixed O$_2$ admixture of 8~ppm, 0.3~W and 500~sccm total He flow.}
	\label{fig:APPJ_H_consumption}
\end{figure}

The good agreement between simulation and experiment allows for the investigation of the most important formation pathways for H. The dominant pathways for production and consumption are shown in \cref{fig:APPJ_H_production,fig:APPJ_H_consumption}, respectively.

H production pathways vary strongly with humidity content. One of the dominant production pathways at low H$_2$O admixtures is via collisions of OH and O (\cref{eq:O_OH}). At higher H$_2$O admixtures, this pathway becomes less relevant. Another important production mechanism for H at low H$_2$O admixtures is the destruction of protonated water clusters and H$_2$O$^+$ via dissociative recombination with electrons 
\begin{align}
e + H^+(H_2O)_n &\rightarrow H + n\cdot H_2O \:,\\
e + H_2O^+ &\rightarrow H + OH \:, \\
e + H_2O^+ &\rightarrow 2H + O \:.
\end{align}
%
%
%
%
%

At higher H$_2$O admixtures, H is predominantly produced by electron impact dissociation of water molecules, reactions between OH and O (\cref{eq:O_OH}), and also to a small extent via dissociation of H$_2$ by electrons 
\begin{align}
e + H_2O &\rightarrow H + OH + e \;, \\
He^* + H_2O &\rightarrow He + H + OH^+ + e \;, \\
e + H_2 &\rightarrow 2H + e \:.
\end{align}
The dissociation of ions and small ion clusters becomes less relevant. This is due to the fact that more water is available for electron impact dissociation. From 50 to 5000~ppm, the H$_2$O content increases by 2 orders of magnitude, while the number densities of cluster ions available for dissociation only increases by a factor of 6. Additionally, high densities of negative ion clusters of the form OH$^-$(H$_2$O)$_n$ are formed, leading to an increasing electro-negativity of the plasma. 


As shown in \cref{fig:APPJ_H_consumption}, the consumption of H is dominated at all humidity contents by recombination at the wall to form H$_2$
\begin{align}
H \rightarrow \frac{1}{2}H_2 \:.
\end{align}
This is reasonable because of the high diffusion coefficient of H atoms, which is inversely proportional to the mass of the particle
\begin{align}
D = \frac{k_BT_g}{m\nu_\text{m}} \:,
\end{align}
where $\nu_\text{m}$ is the collision frequency. Of course the loss of H to the wall is determined by the surface reaction probability $\gamma_\text{H}$ and the return fraction $f_\text{H}$ that is assumed in the simulation. However, as discussed in detail in~\cite{Schroter2018}, these coefficients are usually not measured under similar experimental conditions to those used in this work. Their accuracy should therefore be judged carefully. In this work, as discussed in \cite{Frank2005, Schroter2018}, the H surface recombination probability is assumed to be 0.03 for all conditions.

While at low H$_2$O admixtures, almost all H is consumed by losses to the wall, additional loss mechanisms involving chemical reactions with bulk species become more important at higher H$_2$O contents. In particular, collisions with OH, HO$_2$, O$_2$, and with H itself lead to destruction in the plasma bulk and formation of various short and long-lived species, such as H$_2$O and H$_2$.

\section{Conclusions.}	

In this work, we investigated the formation of atomic oxygen and hydrogen (O and H) in the COST-$\upmu$APPJ in a He-H$_2$O gas mixture using a combination of two-photon absorption laser-induced fluorescence with picosecond temporal resolution (ps-TALIF), and a global model with a He-H$_2$O reaction mechanism. 

The use of ps-TALIF is motivated by the high collisionality in atmospheric pressure plasmas, leading to short lifetimes of the laser-excited states of O and particularly H. With our setup, we were able to directly measure the effective decay rates and also quenching coefficients of these states with H$_2$O molecules. These coefficients have not been studied extensively in the literature and therefore this work provides useful additional data for these coefficients. In the case of O, our value lies between two previously measured values, whereas for H, values from the literature are on average a factor 1.5 larger than our value. In this context, we would like to emphasize the crucial advantage of ps-TALIF not to rely on quenching coefficients in the first place, as the effective lifetime, which is necessary for the determination of absolute species densities, is measured separately for each plasma operating condition.

Experimentally determined O densities reach a maximum of $2.0\times10^{13}$~cm$^{-3}$ at about $3000$~ppm at higher humidity contents in the feed gas ($\ge$ 100 ppm). The trend agrees very well with previous work using vacuum ultra-violet Fourier-transform absorption spectroscopy. Absolute densities are a factor 1.5--2.5, depending on the H$_2$O content, lower compared to these previous results and results obtained by other groups using nanosecond TALIF. 

Crucially, towards very low humidity contents, we find a steep increase of O densities, with a maximum value of $4.3\times10^{13}$~cm$^{-3}$. This has not been observed in previous investigations, which were carried out in closed systems or in artificial atmospheres. We therefore conclude that these high densities are produced from O$_2$ impurities originating from ambient air. This trend has been reproduced using a plug-flow, 0D plasma-chemical kinetics model assuming initial feed gas impurities in the order of ppm. From the simulation we also find that atomic oxygen is mainly produced from O$_2$ impurities (at low H$_2$O content) or via reactions between two OH molecules (at high H$_2$O content), and is consumed via reactions with OH. This also agrees with previous findings. Hence, for reactive species production in plasmas purposely admixed molecules to the feed gas provides a more controllable environment, compared with relying on impurities.

Experimentally determined H densities increase sub-linearly with increasing humidity content up to a density of $6.0\times10^{14}$~cm$^{-3}$. To our knowledge, no comparable data exists yet in the literature. We find good quantitative and qualitative agreement with the plasma-chemistry model results. Using the model further, we find that the main production pathway for H is via dissociation of protonated water clusters and reactions between OH and H (at low water content) and electron impact dissociation of water (at all water content) and the main consumption pathway to be recombination at the reactor wall, which would highly depend on the assumed surface recombination probability in the simulation. 

Since O and H densities show different trends particularly at higher humidity contents (plateau and slight decrease for O, increase for H), changing the humidity content in the plasma yields different O/H ratios. This can be used as a tailoring mechanism for plasma applications. 

\ack
The authors would like to thank Prof. Mark J. Kushner for providing the GlobalKin code and Richard Armitage for assistance with experimental setups.\\
James Dedrick acknowledges financial support from an Australian Government Endeavour Research Fellowship.\\
This work was financially supported by the UK EPSRC (EP/K018388/1 \& EP/H003797/2), and the York-Paris Low Temperature Plasma Collaborative Research Centre.

\section*{References}
\bibliography{AllPapers}
\bibliographystyle{iopart-num}

\end{document}